\documentclass[aps,pra,twocolumn,showkeys,amssymb,floatfix,
longbibliography,superscriptaddress,reprint]{revtex4-1}

\usepackage{graphicx}
\usepackage{amsmath,amsfonts,amssymb}
\usepackage{hyperref}
\usepackage{braket}
\usepackage{xcolor}
\usepackage{tikz}
\usepackage{color}
\usepackage{enumerate}
\usepackage[caption=false]{subfig}
\usepackage{multirow}
\usepackage{qcircuit}
\usepackage[normalem]{ulem}
\usepackage{placeins}
\usepackage{url}
\usepackage{bbm}

\providecommand{\e}[1]{\ensuremath{\times 10^{#1}}}

\newcommand{\appref}[1]{Appendix~\ref{#1}}
\newcommand{\secref}[1]{Sec.~\ref{#1}}
\newcommand{\ssecref}[1]{Section~\ref{#1}}
\newcommand{\figref}[1]{Fig.~\ref{#1}}
\newcommand{\tabref}[1]{Table~\ref{#1}}
\newcommand{\equref}[1]{Eq.~\eqref{#1}}
\newcommand{\equaref}[2]{Eqs.~\eqref{#1} and \eqref{#2}}
\newcommand{\equsref}[2]{Eqs.~\eqref{#1}--\eqref{#2}}

\newcommand{\p}{\partial}

\newcommand{\expect}[1]{\langle #1 \rangle}

\let\baraccent=\=
\renewcommand{\=}[1]{\stackrel{#1}{=}}

\newlength\stextwidth

\newcommand\SI[2]{{#1}\,\mathrm{#2}}
\definecolor{darkgreen}{rgb}{0,0.7,0}

\begin{document}

\title{Testing quantum fault tolerance on small systems}

\author{D. Willsch}
\affiliation{Institute for Advanced Simulation, 
  J\"ulich Supercomputing Centre,\\
  Forschungszentrum J\"ulich, D-52425 J\"ulich, Germany}
\affiliation{RWTH Aachen University, D-52056 Aachen, Germany}
\author{M. Willsch}
\affiliation{Institute for Advanced Simulation, 
  J\"ulich Supercomputing Centre,\\
  Forschungszentrum J\"ulich, D-52425 J\"ulich, Germany}
\affiliation{RWTH Aachen University, D-52056 Aachen, Germany}
\author{F. Jin}
\affiliation{Institute for Advanced Simulation, 
  J\"ulich Supercomputing Centre,\\
  Forschungszentrum J\"ulich, D-52425 J\"ulich, Germany}
\author{H. De Raedt}
\affiliation{Zernike Institute for Advanced Materials,\\
University of Groningen, Nijenborgh 4, NL-9747 AG Groningen, The Netherlands}
\author{K. Michielsen}
\affiliation{Institute for Advanced Simulation, 
  J\"ulich Supercomputing Centre,\\
  Forschungszentrum J\"ulich, D-52425 J\"ulich, Germany}
\affiliation{RWTH Aachen University, D-52056 Aachen, Germany}

\date{\today}

\begin{abstract} We extensively test a recent protocol to demonstrate quantum
  fault tolerance on three systems: (1) a real-time simulation of five spin
  qubits coupled to an environment with two-level defects, (2) a real-time
  simulation of transmon quantum computers, and (3) the 16-qubit processor of
  the IBM Q Experience. In the simulations, the dynamics of the full system is
  obtained by numerically solving the time-dependent Schr\"odinger equation. We
  find that the fault-tolerant scheme provides a systematic way to improve the
  results when the errors are dominated by the inherent control and measurement
  errors present in transmon systems. However, the scheme fails to satisfy the
  criterion for fault tolerance when decoherence effects are dominant.
\end{abstract}


\maketitle

\section{Introduction}\label{sec:intro} A functional universal gate-based
quantum computer requires a very high level of precision in implementing the
quantum gates. In particular when the devices become bigger, it proves difficult
to maintain this high level of qubit control
\cite{sheldon2015singlequbitfidelities, gambetta2015building,
Neil2017GoogleBlueprintQuantumSupremacy, ibmquantumexperience2016,
Willsch2017GateErrorAnalysis} or to satisfy the requirements needed for a
computing device \cite{Michielsen2017BenchmarkingQC}. To overcome these
limitations, the most prominent solution is provided by the theory of
fault-tolerant quantum computation \cite{Shor1996FaultTolerantQC,
Gottesman1998TheoryFTQC, Campbell2017RoadsTowardsFTQC}.

However, despite many experiments on quantum codes
\cite{takita2016demonstration, kelly2015statepreservation9qubits,
chow2014implementingastrand, corcoles2015demonstration, riste2015detecting}, it
has still remained an open question how much a practical application can profit
from a full fault-tolerant protocol. Therefore, Gottesman proposed a test
\cite{Gottesman2016quantumfaulttolerance} that uses four physical qubits to
encode two logical qubits, in combination with a criterion for a successful
demonstration of fault tolerance, requiring that
\begin{center}
  \fbox{\parbox{.8\linewidth}{All encoded
circuits of some representative set perform better than the corresponding bare,
unencoded circuits.}}
\end{center} 
The underlying error-detecting four-qubit
code \cite{Leung1997fourqubitcode, Vaidman1996fourqubitcode,
Grassl1997fourqubitcode} has been implemented with ion-trap qubits
\cite{Linke2016FTIonTrapQubits} and on IBM's five-qubit processor
\cite{Vuillot2017ErrorDetectionIBM, Takita2017faultTolerantStatePreparation,
HarperFlammia2018FaultToleranceInTheIBMQ}.
Each of these experiments reports a successful result, but none explicitly tests
the proposed fault-tolerance criterion.

In this paper, we report on an extensive test of the fault-tolerance criterion
for three complementary systems. System (1) consists of five spin qubits coupled
to an environment at a given temperature. We consider various weak- and strong-coupling 
strengths and various temperatures. This system serves as a general
model to study decoherence \cite{Jin2010approachtoequilibrium,
zhao2016masterequation, Mueller2015twoleveldefects}.  System (2) is an upscaled
version of the real-time circuit-Hamiltonian simulation used in
\cite{Willsch2017GateErrorAnalysis} comprising five transmons and six
resonators.  System (3) is the physical 16-qubit device \texttt{ibmqx5} provided
by IBM \cite{ibmquantumexperience2016}.  We find very good agreement between the
latter two systems for the proper set of optimized gate pulses including
measurement errors. 

The real-time dynamics of both system (1) and (2) are studied by
numerically solving the time-dependent Schr\"odinger equation (TDSE) with
$\hbar=1$, \begin{align} i\frac{\p}{\p t}\ket{\Psi(t)} &= H(t) \ket{\Psi(t)},
  \label{eq:TDSE} \end{align} where $H(t)$ is the time-dependent model
Hamiltonian and $\ket{\Psi(t)}$ represents the state of the device at time $t$.
Note that the computer simulation is a deterministic program that
always produces the same mathematical solution $\ket{\Psi(t)}$, from which we
can compute any physically relevant quantity (such as reduced density matrices
of smaller subsystems with non-unitary dynamics) without the need of sampling
events. A simulation at this level goes, by definition, beyond perturbative
studies, master equations, and assumed Markovianity or completely-positive
trace-preserving maps \cite{Magesan2013ModelingQuantumNoise,
Puzzuoli2014tractablesimulation, Iyer2017smallQCneededforFT}.

We find that, despite the goal of quantum error correction, the fault-tolerant
scheme fails to satisfy the success criterion under the influence of decoherence
errors in system (1). However, our study suggests that fault-tolerant schemes
can systematically improve the performance with respect to the natural control
and measurement errors dominating the transmon systems (2) and (3).

This paper is structured as follows. In
\secref{sec:faulttolerance}, we give a brief overview of the theory of quantum
fault tolerance and the protocol that we study. \ssecref{sec:spinqubits}
contains the results for system (1). In this system, there are no control
errors, allowing us to assess the performance of the fault-tolerant protocol in
the presence of decoherence errors only. In \secref{sec:transmons}, we present
the transmon simulation model, i.e.~system (2). This system allows us to study
the protocol's performance under inherent control and measurement errors.
Subsequently in \secref{sec:ibmq}, we present experimental results for system
(3). This section also contains a comparison with systems (1) and (2), showing
that IBM's transmon qubits are not dominantly affected by decoherence errors and
can thus benefit from the fault-tolerant protocol. Finally, conclusions from our
study of all three systems are given in \secref{sec:discussion}.

\section{Fault tolerance}\label{sec:faulttolerance}

In the framework of quantum fault tolerance, logical qubits are encoded in
multiple physical qubits to allow for the detection and correction of errors.
This concept inevitably relies on a mathematical model for the errors that are
supposed to happen in a physical quantum processor. Simple versions of these
models are based on discrete, uncorrelated single-qubit errors or the
possibility to describe the errors within the quantum operations formalism
\cite{NielsenChuang}, while more sophisticated studies consider non-Markovian
errors in a general Hamiltonian framework
\cite{Terhal2005ftqcForLocalNonmarkovianNoise, aliferis2006extendedrectangles,
  aliferis2007FTQCwithLeakage, aharonov2008thresholdtheorem,
ng2009FTQCversusGaussianNoise}. The results of these studies are so-called
threshold theorems, stating that as long as a certain parameter in the model is
below a certain threshold, arbitrarily long quantum computation is possible.

However, as these threshold theorems are only valid within the
mathematical model for the errors, it is unclear whether a particular quantum
error-correcting scheme is beneficial in an actual application. For instance,
the thresholds are usually expressed in terms of the diamond norm
\cite{kitaev1997diamondnorm}, which is experimentally inaccessible. Although
progress has been made to relate this quantity to the average gate fidelity
\cite{Sanders2016ThresholdTheorem, Kueng2016ComparingExperimentsToThreshold},
recent studies have demonstrated that this fidelity, too, cannot be measured in
a physical quantum information processor
\cite{proctor2017RandomizedBenchmarking}. In fact, it was shown in two
independent studies that none of these error metrics can reliably predict
the performance of quantum gates in a practical application
\cite{Willsch2017GateErrorAnalysis, Iyer2017smallQCneededforFT}.

The fault-tolerant scheme that we test in this study was explicitly designed to
apply to small quantum computers \cite{Gottesman2016quantumfaulttolerance}. It
replaces a bare two-qubit circuit with an encoded four-qubit circuit and an
additional ancilla qubit. In this paper, the term \emph{circuit} is defined to
include both an initial-state preparation and a sequence of gates. In
particular, we consider the initial states
$\ket{00},\ket{0+}=\ket{00}+\ket{01},$ and $\ket{\Phi^+}=\ket{00}+\ket{11}$ (up
to normalization). In the encoded circuits, these states are represented by
entangled four-qubit states (see \appref{app:code} for their definitions and
preparation circuits). Along with the encoding of states, there is a set of
encoded gates to build a quantum circuit. In the present case, this set is given
by $\left\{ \mathrm X1,\mathrm X2,\mathrm Z1,\mathrm Z2,\mathrm{HHS},\mathrm{CZ}
\right\}$, where $\mathrm X1$ and $\mathrm X2$ denote bit-flip gates, $\mathrm
Z1$ and $\mathrm Z2$ denote sign-flip gates, $\mathrm{HHS}$ denotes the Hadamard
gate on each qubit followed by swapping the qubits, and $\mathrm{CZ}$ denotes
the controlled-phase gate \cite{NielsenChuang}. A full specification of how all
bare and encoded circuits are implemented in the fault-tolerant scheme is given
in \tabref{tab:states} and \tabref{tab:gates} in \appref{app:code}.

The aim is to compare the performance of a bare circuit with that of an encoded
circuit for a representative set of circuits. To find such a set, we applied the
procedure suggested in \cite{Gottesman2016quantumfaulttolerance} for the maximum
circuit length $T=10$, the repetition parameter $\mathrm{RP}=6$, and the
periodicity $P=3$, yielding 465 circuits. In this paper, we focus on the results
for a selection of 15 circuits (see \tabref{tab:circuits}) that we consider
representative of the performance of all 465 tested circuits
(cf.~\appref{app:fullcircuits}).

\begin{table}
  \caption{\label{tab:circuits} List of the selected 15 circuits to illustrate
  the difference between bare and encoded versions (see
  \appref{app:fullcircuits} for a list of all 465 tested circuits). The first
  column contains sets of three circuit IDs labeling the circuits in the second
  column, which consist of particular sets of gates operating on three initial
  states $\ket{i}\in\left( \ket{00},\ket{0+},\ket{\Phi^+} \right)$, enumerated
in this order.}
\begin{ruledtabular}
\begin{tabular}{@{}rl@{}}
  ID & Circuit \\
  \colrule
   0-2 & $\ket{i}$ \\
   240-242 & X1 X1 X1 X1 X1 $\ket{i}$ \\
   216-218 & CZ CZ CZ CZ CZ $\ket{i}$ \\
   171-173 & CZ X1 X2 Z1 Z1 X1 X1 Z1 Z1 Z2 $\ket{i}$ \\
   270-272 & HHS CZ HHS CZ HHS CZ HHS CZ HHS CZ $\ket{i}$ \\
\end{tabular}
\end{ruledtabular}
\end{table}

Evaluating the performance of the circuits is done as follows. For the bare
versions, a final measurement of the qubits produces a distribution
$p_{q_3q_4}^{\text{bare}}$ of two-bit strings $q_3q_4$. For the encoded versions,
the same measurement produces a distribution of five-bit strings $q_0q_1q_2q_3q_4$.
The encoding scheme then dictates that if the ancilla qubit $q_0$ is 1 or if the
bit string $q_1q_2q_3q_4$ includes an odd number of 1's (meaning that it does not
correspond to an encoded basis state \cite{Gottesman2016quantumfaulttolerance}),
it is discarded. The ratio of bit strings that are not discarded is called
the \emph{postselection} (PS) ratio $r$. These selected bit strings then
constitute a new distribution $p_{q_3q_4}^{\text{enc}}$, normalized by the PS
ratio $r$. Both bare and encoded distributions can be compared to the
theoretical distribution $p_{q_3q_4}^{\text{theory}}$ that an ideal gate-based
quantum computer produces. The appropriate measure to compare these
distributions is the statistical distance \cite{Sanders2016ThresholdTheorem}
\begin{align}
  \label{eq:Dbare}
  D_{\text{bare}} &= \frac{1}{2} \sum\limits_{q_3q_4}^{}
  \left|p_{q_3q_4}^{\text{bare}} - p_{q_3q_4}^{\text{theory}} \right|,\\
  \label{eq:Denc}
  D_{\text{enc}} &= \frac{1}{2} \sum\limits_{q_3q_4}^{}
  \left|p_{q_3q_4}^{\text{enc}} - p_{q_3q_4}^{\text{theory}} \right|.
\end{align}
In terms of these quantities, Gottesman's success criterion for fault tolerance
is fulfilled if $D_{\text{enc}}<D_{\text{bare}}$ for all circuits under
investigation.

Mathematical motivations suggesting a better performance of the encoded circuits
are (1) the added redundancy in combination with postselection and (2) the fact
that an encoded circuit needs two-qubit gates exclusively for the initial-state
preparation. However, only practical tests such as the one performed in this paper
can tell whether fault-tolerant schemes can improve the performance.

\section{Spin qubits coupled to an environment}\label{sec:spinqubits}

System (1) consists of $5+N_E$ two-level systems. The subsystem with the first
five two-level systems represents the spin qubits of the quantum computer,
and the remaining $N_E$ two-level systems constitute the environment. This model
is motivated by the experimental observation that in recent superconducting
quantum processors, two-level systems
formed by material defects constitute a major source of decoherence caused by
the environment \cite{Mueller2015twoleveldefects,
barendsMartinis2013xmoncoherence, Wang2015dielectricloss}.

\begin{figure}
  \includegraphics[width=.6\linewidth]{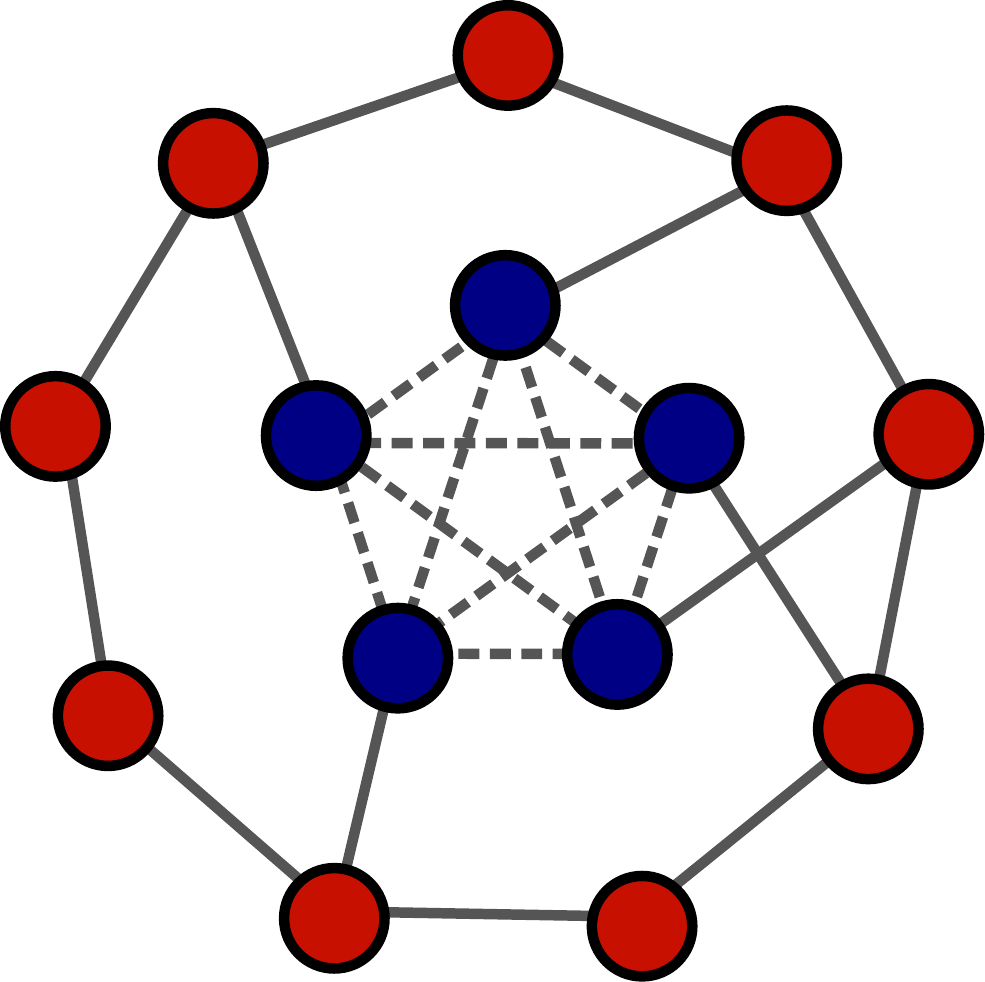}
  \caption{\label{fig:system} (Color online) Schematic representation of the
  system of five spin qubits (blue) coupled to an environment (red), described
  by the model Hamiltonian given in \equsref{eq:H}{eq:HQE}. The five qubits
  representing the quantum computer have a tunable all-to-all connectivity
  (dashed lines).  The two-level systems in the environment form a ring with an
  always-on coupling between nearest neighbors and to the qubits of the quantum
  computer (solid lines).  The latter is controlled by the coupling strength
$\lambda$.}
\end{figure}

We consider the system depicted in \figref{fig:system}. The five qubits have an
all-to-all coupling. Each qubit is connected to one two-level system in the
environment, which is represented by spins organized on a ring.  The Hamiltonian
describing the whole system reads
\begin{align}
  \label{eq:H}
  H &= H_Q + H_E + \lambda H_{QE}
\end{align}
where the Hamiltonians $H_Q$, $H_E$, and $H_{QE}$ describe the quantum computer,
the environment, and the interaction between both, respectively. The parameter
$\lambda$ controls the coupling strength between the quantum computer and the
environment. The Hamiltonians $H_Q$, $H_E$, and $H_{QE}$ given in \equref{eq:H}
read
\begin{align}
  \label{eq:HQ}
  H_Q &= - \sum\limits_{n=0}^{4} \sum\limits_{\alpha=x,z} h_n^\alpha
  \sigma_n^\alpha -\sum\limits_{n,m=0}^4 G_{nm}^x \sigma_n^x \sigma_m^x ,\\
  \label{eq:HE}
  H_E &= -\sum\limits_{n=5}^{N_E+4} \sum\limits_{\alpha=x,y,z} J_n^\alpha
  \sigma_n^\alpha \sigma_{n+1}^\alpha ,\\
  \label{eq:HQE}
  H_{QE} &= -\sum\limits_{n=0}^{4}\sum\limits_{\alpha=x,y,z} K_{nj_n}^\alpha
  \sigma_n^\alpha \sigma_{j_n}^\alpha,
\end{align}
where $\sigma_n^\alpha$ for $\alpha=x,y,z$ denote the Pauli matrices for qubit
$n$. Each qubit $n\in\left\{ 0,\ldots,4 \right\}$ is connected to a randomly
chosen qubit $j_n\in\left\{ 5,\ldots,N_E+4 \right\}$ in the environment (all
$j_n$ are different) with a random coupling strength $\lambda
|K_{nj_n}^\alpha|\approx \lambda\times\SI 2 {GHz}$, tunable through the
parameter $\lambda$. In the environment Hamiltonian $H_E$, the couplings
$J_n^\alpha$ are chosen randomly from $\left[ -J,J \right]$ for $J=\SI 2
{GHz}$.

Implementing the quantum gates through piecewise constant
parameters in \equref{eq:HQ} eliminates any
control and measurement errors.  Therefore, with this implementation, we can
exclusively study the effect of decoherence errors because the only source of
errors is the interaction between the qubits and the environment.  The
comparison with the results of systems (2) and (3) then allows us to
understand the difference between decoherence errors and control or measurement
errors when using a fault-tolerant protocol.  In \appref{app:spinqubits}, we
give the full specification  of the parameters $h_n^\alpha$ and $G_{nm}^x$ that
enter in $H_Q$ (see \equref{eq:HQ}).  In the absence of coupling to the
environment, the whole system evolves in time like an ideal quantum computer.
Running all quantum circuits on both system (1) for $\lambda=0$ and the J\"ulich
universal quantum computer simulator~\cite{DeRaedt2017MassivelyParallel}  yields
identical results, validating the correct implementation of the quantum
gates.

We solve the TDSE given in \equref{eq:TDSE} with the piecewise time-independent
Hamiltonian given in \equsref{eq:H}{eq:HQE} to machine precision by means of the
Chebyshev polynomial representation of $\exp(-itH)$
\cite{talezerkosloff1984chebyshev,
dobrovitski2003chebyshev,DeRaedt2017relaxation}. The environment is prepared at
an inverse temperature $\beta$ using the random-state technology
\cite{HamsDeRaedt2000RandomStateTechnology, DeRaedt2017relaxation}. 

To understand how $\lambda$ affects the qubit coherence, we estimate the
decoherence time $T_2$ by preparing the qubit along the positive $x$ axis,
evolving it in the presence of the environment, and fitting a damped oscillation
to the decay of its projection on the $x$ axis; see
\cite{deraedt2012rabioscillations} for more details on this procedure.
These experiments are performed at inverse temperature $\beta=0$ to
produce the worst-case decoherence times. We find that
$T_2^{\lambda}\approx \SI{3.7}{ns}/\lambda^2$ (data not shown) and,
specifically,
$T_2^{\lambda=0.1}\approx\SI{370}{ns}$ and
$T_2^{\lambda=0.01}\approx\SI{4\times10^4}{ns}$. In particular, the decoherence
time $T_2^{\lambda=0.01}$ is much larger than the time needed to execute a
quantum circuit in this model (cf.~\tabref{tab:envgateparameters} in
\appref{app:spinqubits}), which supports the interpretation of $\lambda=0.01$
representing a very weak coupling between the ideal quantum computer and the
environment.

\begin{figure}
  \includegraphics[width=\linewidth]{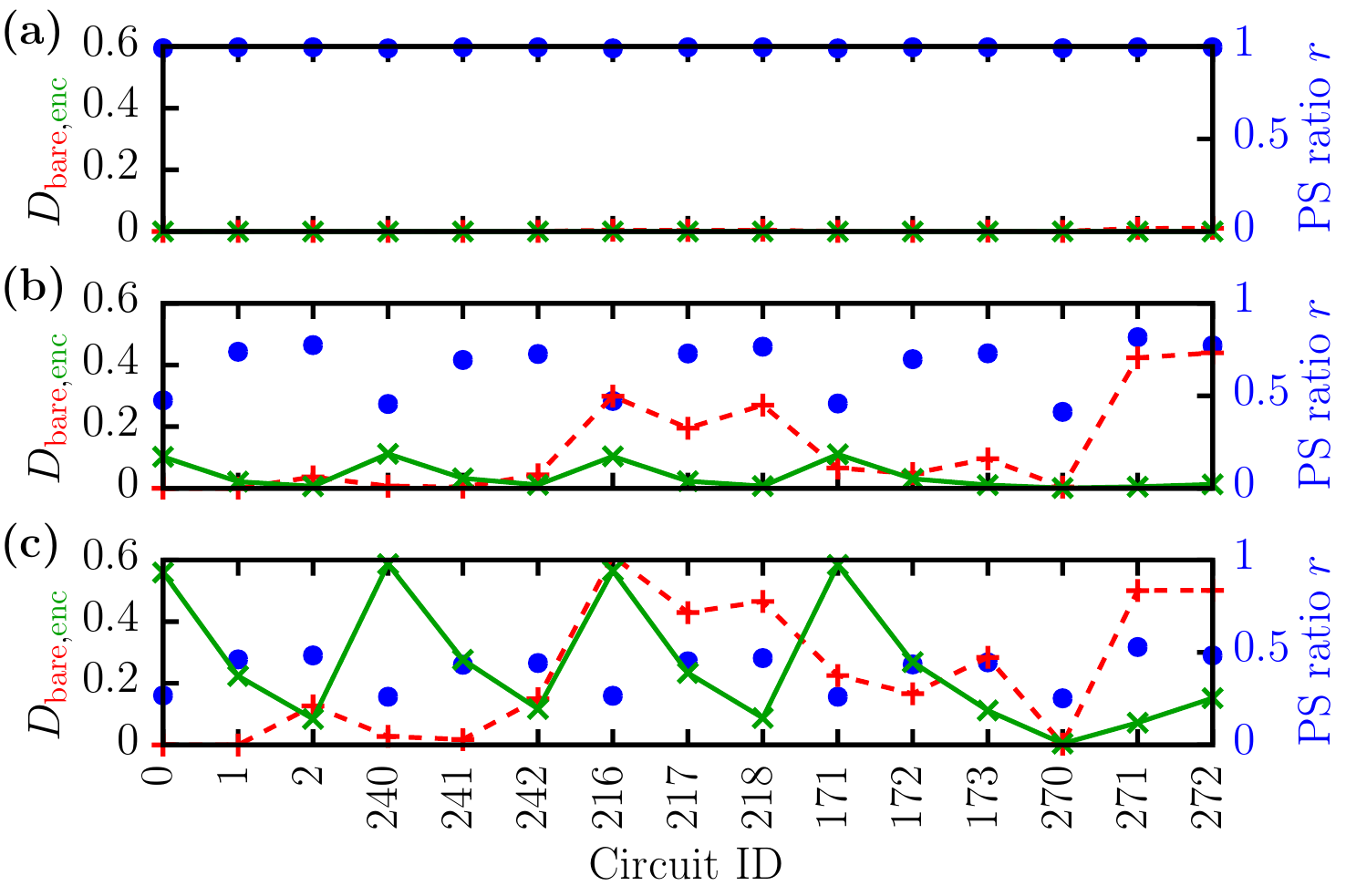}
  \caption{\label{circuitslambda} (Color online) Test of the fault-tolerance
  criterion in system (1) for different coupling strengths (a) $\lambda=0.01$,
  (b) $\lambda=0.1$, and (c) $\lambda=0.2$ between the qubits and the environment.
  Shown are the statistical distances to the ideal result for the selected bare
  (dashed red line) and encoded (solid green line) circuits as defined in
  \equaref{eq:Dbare}{eq:Denc}, and the postselection ratios (blue dots).
  All plotted quantities are dimensionless.
  The simulations were done for inverse temperature $\beta=1$ and environment
$N_E=20$. Lines connecting the data points are guides to the eye.}
\end{figure}

In \figref{circuitslambda}, we present results for the statistical
distances $D_{\text{bare}}$ and $D_{\text{enc}}$ (see
\equaref{eq:Dbare}{eq:Denc}) and the PS ratio $r$ for the circuits listed in
\tabref{tab:circuits}. The three cases shown in \figref{circuitslambda}(a), (b),
and (c) are representative of the transition from very weak coupling
$\lambda=0.01$ to strong coupling $\lambda=0.2$ between the qubits and the
environment. For the weakest coupling (see \figref{circuitslambda}(a)), the
statistical distances for both bare and encoded circuits are nearly zero, and
the postselection ratios $r\approx1$.  This shows that in this case, both bare
and encoded versions perform almost perfectly (i.e.~both produce the ideal
result used in \equaref{eq:Dbare}{eq:Denc}). This observation also demonstrates
the correct implementation of the quantum computer by means of the model defined
by \equsref{eq:H}{eq:HQE}. 

Increasing the coupling strength $\lambda$ leads to a stronger
influence of decoherence errors on the operation of the quantum computer.
Accordingly, in \figref{circuitslambda}(b) and (c), it can be seen that the
statistical distances of both bare and encoded circuits increase. Interestingly,
one can always find circuits for which the bare version outperforms the encoded
version. In particular, every third circuit starting from circuit ID 0 shows a
strong increase in $D_{\text{enc}}$. These circuits correspond to the encoding
of the state $\ket{00}$. The physical reason behind the sensitivity of these
circuits is that the encoding circuit for $\ket{00}$ includes the largest number
of two-qubit gates (see \tabref{tab:states} in \appref{app:code}). These
two-qubit gates typically take a longer time to execute than single-qubit gates
\cite{chow2014implementingastrand, sheldon2016procedure,
Linke2016FTIonTrapQubits} (see also \tabref{tab:envgateparameters} in
\appref{app:spinqubits}). Hence, the entangling two-qubit gate is the most
sensitive gate even when no control errors, but only decoherence errors, are
present. The only exception is the circuit with ID 270, which always yields
$D_{\text{bare}}\approx D_{\text{enc}}\approx0$ (see \figref{circuitslambda}).
The reason is that the execution time of this circuit is so long that the
interaction with the environment leads to a uniform distribution of all
five-qubit states in the quantum computer, which accidentally matches the ideal
output distribution (cf.~\tabref{tab:circuits}).

\begin{table}
  \caption{\label{tab:circuitslambda} Percentage $P$ of the circuits from
  \tabref{tab:circuits} for which the encoded version performs better than the bare
  version, as a function of the coupling strength $\lambda$ between the qubits
  and the environment. The coupling strengths range from very weak to strong
  coupling. The simulations were done for inverse temperature $\beta=1$ and
  environment size $N_E=20$.}
\begin{ruledtabular}
\begin{tabular}{@{}lccccccccc@{}}
  $\lambda$ & 0.01 & 0.025 & 0.05 & 0.075 & 0.1  & 0.125 & 0.15 & 0.175 & 0.2 \\
  $P$ & 80\% & 87\%  & 73\% & 73\%  & 67\% & 67\%  & 53\% & 53\%  & 53\% \\
\end{tabular}
\end{ruledtabular}
\end{table}

A summary of the performance for various intermediate coupling
strengths $\lambda\in\left[ 0.01,0.2 \right]$ is given in
\tabref{tab:circuitslambda}. Interestingly, the percentage $P$ of encoded
circuits performing better than the bare circuits is not a monotonous function
of $\lambda$. For instance, the largest value of $P$ is found at $\lambda=0.025$
instead of $\lambda=0.01$. However, for such a weak coupling, both bare and
encoded circuits perform nearly perfectly (cf.~\figref{circuitslambda}(a)).

In addition to the results shown in \figref{circuitslambda} and
\tabref{tab:circuitslambda}, we have studied the performance of the circuits for
different environment sizes $N_E\in\{5,20,27\}$ (see \figref{circuitsNE} in
\appref{app:spinqubits}) and inverse temperatures $\beta\in\{0,1,5\}$ (see
\figref{circuitsbeta} in \appref{app:spinqubits}), each of which yields results
with the characteristic features resembling those in
\figref{circuitslambda}(b).  This means that in all analyzed regimes, there are
always some encoded circuits that perform worse than their bare equivalents. In
other words, we did not find any case that passes the fault-tolerance test.

One may ask whether this result violates the threshold theorems proven in
\cite{Terhal2005ftqcForLocalNonmarkovianNoise, aliferis2006extendedrectangles,
aliferis2007FTQCwithLeakage, aharonov2008thresholdtheorem}, which obviously
consider a Hamiltonian similar to \equsref{eq:H}{eq:HQE}. The answer is that in
the threshold theorems, the required value of $\lambda$ is still orders-of-magnitude 
smaller than the ones we studied. Yet, already for $\lambda=0.01$ (see
\figref{circuitslambda}(a)), both bare and encoded circuits perform almost
perfectly and encoding still makes the result worse in some cases. We conclude
that using a fault-tolerant protocol such as the one suggested in
\cite{Gottesman2016quantumfaulttolerance} to overcome errors in a system
dominated by decoherence errors from two-level defects is not necessarily
helpful.

\section{Transmon simulation}\label{sec:transmons}

\begin{figure}
  \includegraphics[width=\linewidth]{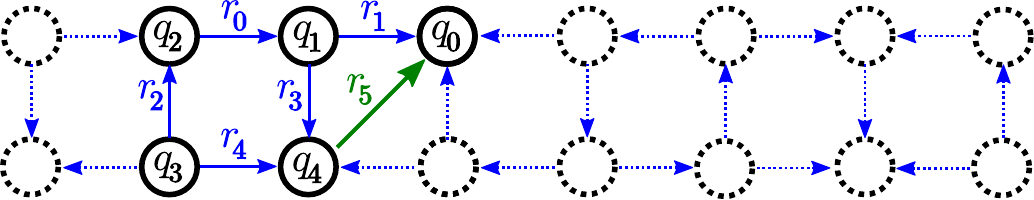}
  \caption{\label{topology} (Color online) Schematic image showing the five
  transmon qubits and six resonators described by the Hamiltonian given in
  \equsref{eq:modelhamiltonian}{eq:modelhamiltonianres}. The system represents a
  subset of the 16-qubit device \texttt{ibmqx5} \cite{ibmquantumexperience2016}
  with an additional resonator $r_5$ to enable the implementation of all bare
  and encoded circuits. Without this resonator, the encoded circuits with
  initial state $\ket{00}$ cannot be fault-tolerantly implemented
  \cite{Gottesman2016quantumfaulttolerance}.}
\end{figure}

System (2) is defined by the circuit Hamiltonian for $N_{\text{tr}}=5$
superconducting transmon qubits coupled by $N_{\text{res}}=6$ transmission-line
resonators \cite{koch2007transmon, blais2004circuitqed}, a system that can be
used to model IBM's publicly accessible quantum processors
\cite{Willsch2017GateErrorAnalysis, ibmquantumexperience2016}. The simulated
system is schematically shown in \figref{topology} as a subset of the 16-qubit
device \texttt{ibmqx5} \cite{ibmquantumexperience2016}.

The full Hamiltonian used in the transmon simulation reads
\begin{align}
  \label{eq:modelhamiltonian}
  H &= H_{\text{tr}} + H_{\text{res}},\\
  \label{eq:modelhamiltoniantr}
  H_{\text{tr}} &= \sum\limits_{i} \left[ 4 E_{Ci}(\hat n_i - n_{gi}(t))^2 -
  E_{Ji} \cos\hat\varphi_i \right],\\
  \label{eq:modelhamiltonianres}
  H_{\text{res}} &= \sum\limits_{r} \Omega_r\hat a_r^\dagger\hat a_r +
  \sum\limits_{r,i} G_{ri} \hat n_i(\hat a_r + \hat a_r^\dagger),
\end{align}
where $i=0,\ldots,N_{\text{tr}}-1$ enumerates the transmon qubits with
capacitive energies $E_{Ci}$, Josephson energies $E_{Ji}$, number operators
$\hat{n}_i$, and superconducting phase operators $\hat\varphi_i$. The resonators
are labeled by $r=0,\ldots,N_{\text{res}}-1$ and are described by their raising and
lowering operators $\hat a_r^\dagger$ and $\hat a_r$, respectively. Their
frequencies are given by $\Omega_r$ and the capacitive coupling strength between
transmon $i$ and resonator $r$ is denoted by $G_{ri}$. Quantum gates on the
transmons are implemented through microwave voltage pulses represented by
$n_{gi}(t)$ \cite{Willsch2017GateErrorAnalysis}. A specification of all device
parameters and pulse shapes is given in \appref{app:transmons}.

We simulate the transmon computer model defined in
\equsref{eq:modelhamiltonian}{eq:modelhamiltonianres} by solving the TDSE given
in \equref{eq:TDSE} with the time-dependent Hamiltonian in
\equref{eq:modelhamiltonian} using a second-order Suzuki-Trotter product-formula
algorithm \cite{deraedt1987productformula, deraedt2004computational,
Willsch2017GateErrorAnalysis} with time step $\tau=\SI{0.001}{ns}$. 
The simulation includes as many higher levels in the transmons and the
resonators as necessary to describe the dynamics of the system accurately (see
\appref{app:transmons} for more information).
The device parameters in \equsref{eq:modelhamiltonian}{eq:modelhamiltonianres} and
optimized gate pulses $n_{gi}(t)$ are chosen such that they represent a subset
of five transmons and five resonators from the 16-qubit device \texttt{ibmqx5}
\cite{ibmquantumexperience2016, McKay2016VZgate, sheldon2016procedure}.
Additionally, a sixth resonator $r_5$ is included in the model (see
\figref{topology}) to extend the connectivity such that all circuits of the
fault-tolerant scheme can be implemented. The additional resonator
solves the problem faced in \cite{Vuillot2017ErrorDetectionIBM}, where the
original fault-tolerant encoding could not be implemented and an alternative
encoding was used which, although fault-tolerant in theory, did not pass the
fault-tolerance test on the IBM device.

\begin{figure}
  \includegraphics[width=\linewidth]{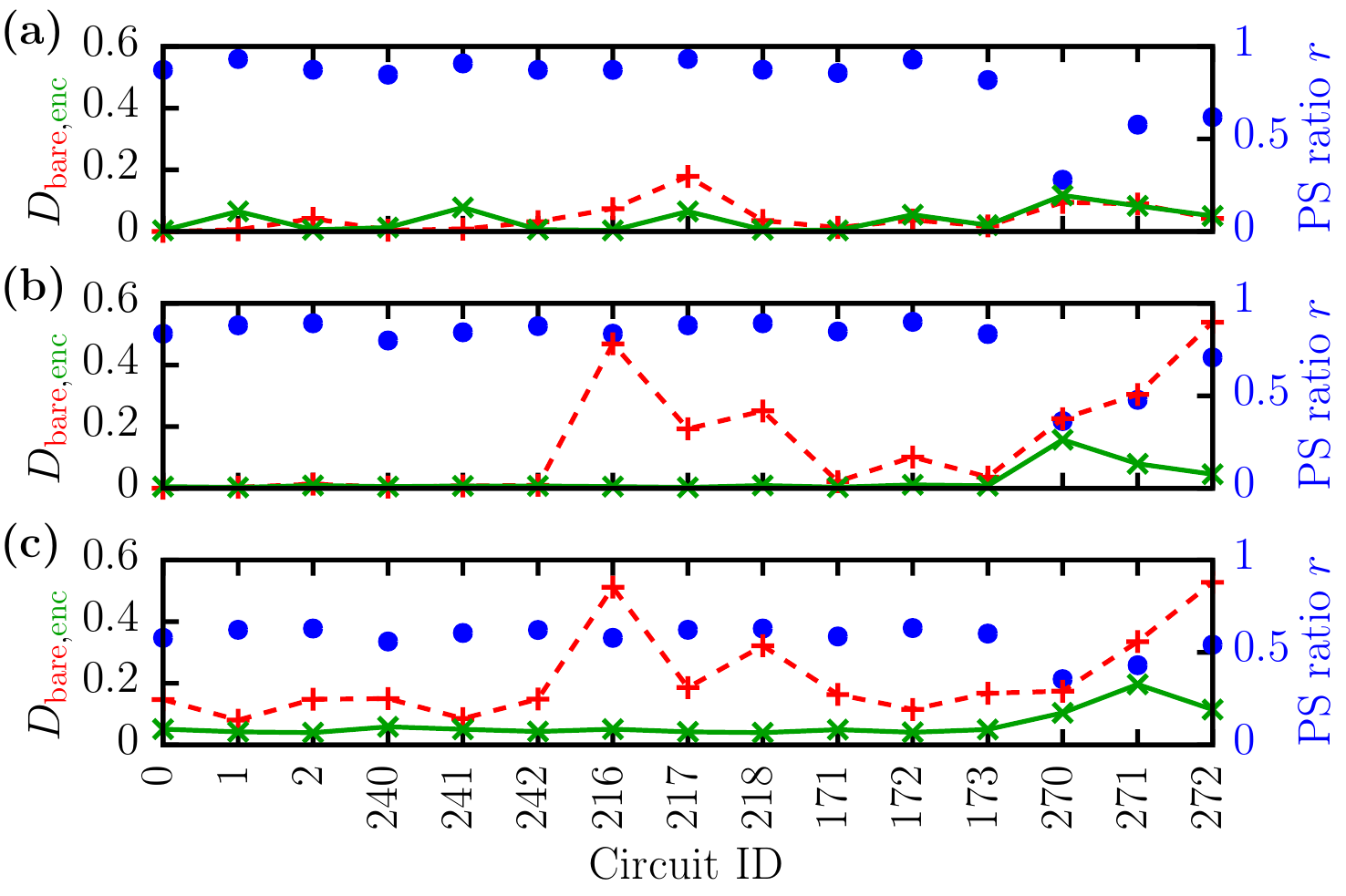}
  \caption{\label{ibm5ed} (Color online) Test of the fault-tolerance criterion
  in system (2), i.e., the real-time transmon simulation for different optimized
  gate sets (a) without frequency tuning, (b) with frequency tuning, and (c)
  with frequency tuning and measurement error $p=0.08$. Shown are the
  statistical distances to the ideal result for the selected bare (dashed red
  line) and encoded (solid green line) circuits as defined in
  \equaref{eq:Dbare}{eq:Denc}, and the postselection ratios (blue dots). 
  All plotted quantities are dimensionless.
  Lines connecting the data points are guides to the eye.}
\end{figure}

The results of the fault-tolerance test are shown in \figref{ibm5ed} for two
different gate sets. Both gate sets use Gaussian microwave pulses
driven at a certain drive frequency $f$ to implement the quantum gates (see
\equref{eq:singlequbitpulse}, \tabref{tab:xgateparameters}, and
\tabref{tab:cnotgateparameters} in \appref{app:transmons} for the individual
parameters resulting from the pulse optimization). For the first gate set, this
drive frequency was set to the respective qubit frequency for each qubit. As can
be seen in \figref{ibm5ed}(a), the performance is equally good for both bare and
encoded circuits. The fault-tolerance criterion $D_{\text{enc}}<D_{\text{bare}}$
is not satisfied. 

The second gate set has been obtained by additionally optimizing the drive
frequencies of the microwave pulses. This means that the drive frequencies are
slightly detuned from the qubit frequencies such that the gate fidelities are
slightly better on average (compare \cite{gambetta2010dragtheory} and
\tabref{tab:gatemetrics} in \appref{app:transmons}; note, however, that better
fidelities do not always imply better gates
\cite{Willsch2017GateErrorAnalysis}). Unlike the first gate set, the second gate
set shows nearly perfect performance for all the encoded circuits (see
\figref{ibm5ed}(b)), suggesting that a fault-tolerant implementation can profit
more from reduced control errors than a bare implementation. In particular, by
examining the numerical results used for \figref{ibm5ed}(b), we find that the
fault-tolerance criterion is satisfied for all circuits but the one with ID 0
(corresponding to $\ket{00}$; see \tabref{tab:circuits}). 
This is the only
circuit for which the bare version does not 
require any pulses and, obviously, applying no pulse is bound to perform better than
applying the preparation pulses to encode $\ket{00}$. Therefore, in the absence
of additional measurement errors, this exception is reasonable.

To assess the effect expected due to measurement errors, we model an additional
error for each qubit such that with probability $p$, a measured bit 0 is
erroneously counted as 1, and vice versa. As shown in \figref{ibm5ed}(c) for the
case $p=0.08$, the fault-tolerance test is passed for all circuits. Thus, in
addition to the natural unitary errors inherently included in the real-time
transmon simulation (cf.~\cite{Willsch2017GateErrorAnalysis}), the presence of
measurement errors is essential to fulfill the fault-tolerance criterion.

\begin{figure}
  \includegraphics[width=\linewidth]{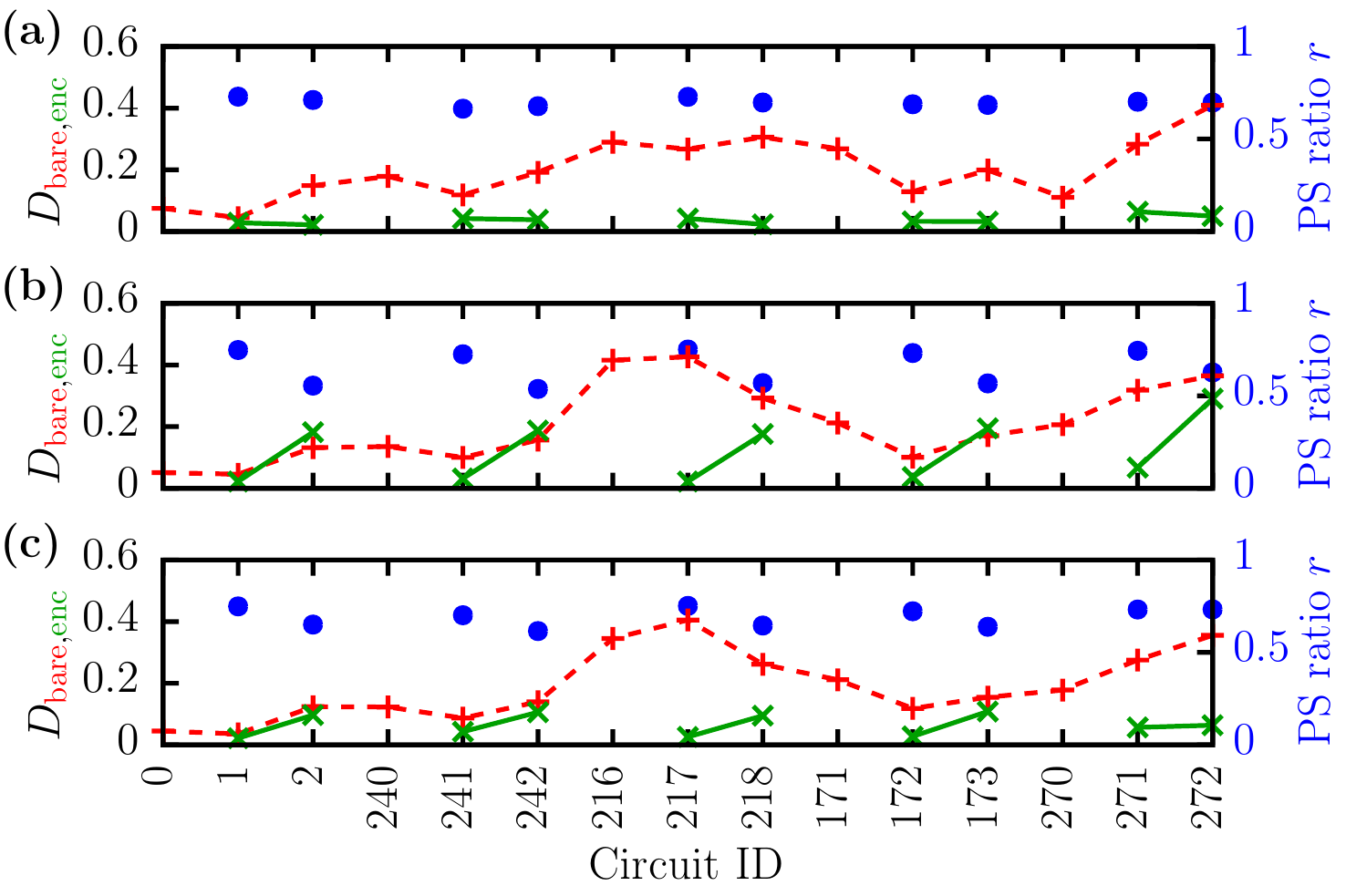}
  \caption{\label{ibmqx5} (Color online) Test of the fault-tolerance criterion
  in system (3), i.e., the 16-qubit device \texttt{ibmqx5} using the qubits
  $(Q_4,Q_3,Q_2,Q_{15},Q_{14})$ on (a) April 3, 2018, (b) April 9, 2018, and (c)
  April 19, 2018. Shown are the statistical distances to the ideal result for
  the selected bare (dashed red line) and encoded (solid green line) circuits as
  defined in \equaref{eq:Dbare}{eq:Denc}, and the postselection ratios (blue
  dots).
  All plotted quantities are dimensionless.
  Only the circuits that could be mapped on the topology were run on the
  real device. Lines connecting the data points are guides to the eye.}
\end{figure}

\section{Physical transmon device}\label{sec:ibmq}

System (3) is used to test the fault-tolerance criterion by utilizing the
16-qubit device \texttt{ibmqx5} provided by IBM \cite{ibmquantumexperience2016}.
Using the qubit mapping $q_0q_1q_2q_3q_4\mapsto Q_4Q_3Q_2Q_{15}Q_{14}$, this
device provides the correct connectivity to run all circuits except for the
encoded version of the circuits with initial state $\ket{00}$ (a problem which
was solved in system (2) by including the additional resonator $r_5$, see
\figref{topology}). 

The results for 15 out of the 465 tested circuits are shown in \figref{ibmqx5}
for three different calibrations. We observe that the performance of the device
varies for different calibrations. For instance, the experiment on April 9, 2018
shown in \figref{ibmqx5}(b) failed the fault-tolerance test. However, in general
many runs passed the test for all circuits (see also \figref{ibmqx5full} in
\appref{app:fullcircuits} for the full set of circuits).

As system (2), discussed in \secref{sec:transmons}, was designed to
simulate a transmon processor such as \texttt{ibmqx5}, it is of course tempting to
set the results in relation to the experimental observations presented in
\figref{ibmqx5}. The first set of gate pulses used for \figref{ibm5ed}(a), where
the drive frequencies were not optimized but set to the qubit frequencies, shows
a circuit performance that differs from the results shown in \figref{ibmqx5}. In
particular, the fault-tolerance test fails. However, the second gate set used
for \figref{ibm5ed}(b) yields a positive result for all circuits that often
passed the test on \texttt{ibmqx5} (see \figref{ibmqx5}(a) and (c)). This
suggests that the pulses used on IBM's processor also utilize slightly detuned
drive frequencies.

Note that the individual circuit performance of system (3) depends a lot on the
pulse parameters found in the calibration procedure. For instance, in
\figref{ibmqx5}(b) and (c), the encoded circuits with initial state
$\ket{\Phi^+}$ always perform slightly worse than the encoded circuits with
initial state $\ket{0+}$. We examined the gate errors reported by IBM for the
corresponding two-qubit gates, finding that they reflect this observation on
April 9 and April 19. The reason that we cannot observe this feature in
\figref{ibm5ed} for system (2) is that our pulse-optimization procedure produces
slightly more reliable pulse parameters whose two-qubit error rates do not
spread as much and also do not differ between runs on separate days
(cf.~\tabref{tab:gatemetrics} in \appref{app:transmons}).

The best agreement between simulation and experiment is achieved when an
additional measurement error is taken into account (see \figref{ibm5ed}(c)). In
particular, the fault-tolerance criterion is then also satisfied for every
encoded circuit that could not be run on \texttt{ibmqx5} (corresponding to the
circuit IDs 0, 240, 216, 171, and 270).  This suggests that the positive result
for the fault-tolerance test may also be observed if the device's connectivity
is extended to support the complete set of circuits, as was done in the
simulation (see \figref{topology}). 

A direct comparison to system (1), i.e., the system of spin qubits coupled to an
environment discussed in \secref{sec:spinqubits}, yields another interesting
conclusion.  Clearly, the performance of the tested circuits shown in
\figref{circuitslambda} differs largely from the results shown in
\figref{ibmqx5} in that the fault-tolerance criterion for system (1) was not
satisfied for any of the studied set of parameters. This led to the conclusion
that decoherence errors are difficult to mitigate with the fault-tolerant
scheme. However, \figref{ibmqx5} and \figref{ibmqx5full} (see
\appref{app:fullcircuits}) show that the fault-tolerance criterion can indeed be
achieved in the IBM Q Experience.  Thus we conclude that the errors in IBM's
quantum processors are not dominated by decoherence from material defects.

\section{Discussion}\label{sec:discussion}

We have tested a full fault-tolerant protocol encoding two logical
qubits on three complementary systems, each dominated by a certain type of
errors present in applications. Since these errors can be much more complicated
than those assumed in the design of fault-tolerant protocols, it is by no means
guaranteed that using a fault-tolerant protocol improves the computation.

System (1) is a set of five spin qubits coupled to an environment with various
coupling strengths, sizes, and temperatures. This system suffers only from
decoherence errors that are controlled by the coupling strength. We found that
the fault-tolerance criterion is not satisfied for any set of parameters,
suggesting that dominating decoherence errors are hard to mitigate with a
fault-tolerant scheme.

System (2) is a model system of five transmon qubits and six resonators, in
which the quantum gates are implemented by the same Gaussian microwave pulses
that are also used in experiments \cite{gambetta2012crosstalkSimRB,
McKay2016VZgate, sheldon2016procedure, Willsch2017GateErrorAnalysis}. Lacking an
environment, this system's performance is purely affected by unitary control
errors. We found that for the appropriate set of gate pulses with detuned drive
frequencies, a full fault-tolerant protocol can systematically improve a quantum
computer's performance. In the presence of an additional measurement error, we
showed that the fault-tolerance criterion is satisfied for all circuits under
investigation.

System (3) is a physical implementation of a quantum computer based on transmon
qubits, namely, the device \texttt{ibmqx5} of the IBM Q Experience
\cite{ibmquantumexperience2016}. While the results varied with the day on which
we carried out the experiments, the general observation was that the
fault-tolerance criterion is satisfied for all circuits that could be mapped on
the device topology. Furthermore, by comparing the experimental results with the
simulation results for system (2), we found that this observation still holds
if the device's topology is extended to support the complete set of circuits.  A
comparison with the results for system (1) further suggests that the errors
in IBM's quantum processor are largely control and measurement errors, implying
that the device is well isolated from decoherence due to material defects.

Based on these results, we conclude that the performance of a quantum
computer can be systematically improved with a fault-tolerant protocol, as
long as the errors of the underlying processor are due to control and
measurement errors. However, the use of a fault-tolerant scheme is not
necessarily helpful when decoherence errors are dominant.

\begin{acknowledgments}

We acknowledge use of the IBM Q Experience for this work. The views expressed
are those of the authors and do not reflect the official policy or position of
IBM or the IBM Q Experience team. D.W. is supported by the Initiative and
Networking Fund of the Helmholtz Association through the Strategic Future Field
of Research project ``Scalable solid state quantum computing (ZT-0013).'' The 
authors acknowledge the computing time granted by the JARA-HPC
Vergabegremium and provided on the JARA-HPC Partition part of the supercomputer
JUQUEEN~\cite{JUQUEEN} at the Forschungszentrum J\"ulich.

\end{acknowledgments}

\appendix
\onecolumngrid

\section{Specification of the fault-tolerant scheme}\label{app:code}

The error-detecting code used in the fault-tolerant scheme is the
$[[4,2,2]]$ code, where the logical two-qubit states are defined as
\begin{align}
  \overline{\ket{00}} &= (\ket{0000}+\ket{1111})/\sqrt{2}\label{00L} ,\\
   \overline{\ket{01}} &= (\ket{1100}+\ket{0011})/\sqrt{2} ,\\
   \overline{\ket{10}} &= (\ket{1010}+\ket{0101})/\sqrt{2} ,\\
   \overline{\ket{11}} &= (\ket{0110}+\ket{1001})/\sqrt{2}\label{11L}.
\end{align}
By linear combination, one can derive the encoded versions of the other
two initial states considered in this study,
\begin{align}
   \overline{\ket{0+}} &= (\ket{0000}+\ket{1100}+\ket{0011}+\ket{1111})/2 ,\\
   \overline{\ket{\Phi^+}} &= (\ket{0000}+\ket{0110}+\ket{1001}+\ket{1111})/2.
\end{align}
The physical four-qubit states are enumerated in increasing order as
$q_1q_2q_3q_4$ and an additional qubit $q_0$ is used as the ancilla qubit. Using
this labeling, the logical gates used in the tested circuits map to the physical
gates according to
\begin{align}
  \overline{\mathrm X1} &= X_1 X_3 \label{X1L},\\
   \overline{\mathrm X2} &= X_1 X_2 ,\\
   \overline{\mathrm Z1} &= Z_1 Z_2 ,\\
   \overline{\mathrm Z2} &= Z_1 Z_3 ,\\
   \overline{\mathrm{HHS}} &= H_1 H_2 H_3 H_4,\\
   \overline{\mathrm{CZ}} &= S_1 S_2 S_3 S_4 Z_2 Z_3 \label{CZL},
\end{align}
which can be easily verified by applying the logical gates to the definition
  of the logical states given in \equsref{00L}{11L}.
In \tabref{tab:states} and \tabref{tab:gates}, we give a specification of all
gate sequences used to assemble the bare and encoded versions of the circuits
to test the fault-tolerance criterion.

\begin{table}
  \caption{\label{tab:states} Initial states and the bare and encoded versions
of their preparation circuits.}
\begin{ruledtabular}
  \begin{tabular}{@{}lll@{}}
  State & Bare version & Encoded version \\
  \colrule
  $\ket{00}$
  &
  \Qcircuit @C=1.5em @R=1.5em @!R {
  \lstick{\mathrm{q_3}\ket{0}}&\qw\\
  \lstick{\mathrm{q_4}\ket{0}}&\qw\\&}
  & 
  \Qcircuit @C=.5em @R=.2em @!R {
  \lstick{\mathrm{q_0}\ket{0}}&\qw     &\qw      &\qw      &\qw      &\targ    &\targ    &\meter&\cw\\ 
  \lstick{\mathrm{q_1}\ket{0}}&\qw     &\qw      &\targ    &\qw      &\ctrl{-1}&\qw      &\qw   &\qw\\
  \lstick{\mathrm{q_2}\ket{0}}&\qw     &\targ    &\ctrl{-1}&\qw      &\qw      &\qw      &\qw   &\qw\\
  \lstick{\mathrm{q_3}\ket{0}}&\gate{H}&\ctrl{-1}&\qw      &\ctrl{1} &\qw      &\qw      &\qw   &\qw\\
  \lstick{\mathrm{q_4}\ket{0}}&\qw     &\qw      &\qw      &\targ    &\qw      &\ctrl{-4}&\qw   &\qw\\&}
  \\
  \colrule
  $\ket{0+}$
  & 
  \Qcircuit @C=.5em @R=.5em @!R {
  \lstick{\mathrm{q_3}\ket{0}}&\qw     &\qw\\
  \lstick{\mathrm{q_4}\ket{0}}&\gate{H}&\qw\\&}
  &
  \Qcircuit @C=.5em @R=.2em @!R {
  \lstick{\mathrm{q_1}\ket{0}}&\qw     &\targ    &\qw      &\qw      &\qw\\
  \lstick{\mathrm{q_2}\ket{0}}&\gate{H}&\ctrl{-1}&\qw      &\qw      &\qw\\
  \lstick{\mathrm{q_3}\ket{0}}&\qw     &\qw      &\gate{H} &\ctrl{1} &\qw\\
  \lstick{\mathrm{q_4}\ket{0}}&\qw     &\qw      &\qw      &\targ    &\qw\\&}
  \\
  \colrule
  $\ket{\Phi^+}$
  & 
  \Qcircuit @C=.5em @R=.5em @!R {
  \lstick{\mathrm{q_3}\ket{0}}&\gate{H}&\ctrl{1} &\qw\\
  \lstick{\mathrm{q_4}\ket{0}}&\qw     &\targ    &\qw\\&}
  &
  \Qcircuit @C=.5em @R=.2em @!R {
  \lstick{\mathrm{q_1}\ket{0}}&\qw     &\qw      &\gate{H} &\ctrl{3} &\qw\\
  \lstick{\mathrm{q_2}\ket{0}}&\qw     &\targ    &\qw      &\qw      &\qw\\
  \lstick{\mathrm{q_3}\ket{0}}&\gate{H}&\ctrl{-1}&\qw      &\qw      &\qw\\
  \lstick{\mathrm{q_4}\ket{0}}&\qw     &\qw      &\qw      &\targ    &\qw\\&}
  \\
\end{tabular}
\end{ruledtabular}
\end{table}

\begin{table}
  \caption{\label{tab:gates} Bare and encoded gate elements.}
\begin{ruledtabular}
  \begin{tabular}{@{}lll@{}}
  Gate & Bare version & Encoded version \\
  \colrule
  X1
  & 
  \Qcircuit @C=.5em @R=.5em @!R {
  \lstick{\mathrm{q_3}}&\gate{X} &\qw\\
  \lstick{\mathrm{q_4}}&\qw      &\qw\\&}
  &
  \Qcircuit @C=.5em @R=.2em @!R {
  \lstick{\mathrm{q_1}}&\gate{X} &\qw\\
  \lstick{\mathrm{q_2}}&\qw      &\qw\\
  \lstick{\mathrm{q_3}}&\gate{X} &\qw\\
  \lstick{\mathrm{q_4}}&\qw      &\qw\\&}
  \\
  \colrule
  X2
  & 
  \Qcircuit @C=.5em @R=.5em @!R {
  \lstick{\mathrm{q_3}}&\qw      &\qw\\
  \lstick{\mathrm{q_4}}&\gate{X} &\qw\\&}
  &
  \Qcircuit @C=.5em @R=.2em @!R {
  \lstick{\mathrm{q_1}}&\gate{X} &\qw\\
  \lstick{\mathrm{q_2}}&\gate{X} &\qw\\
  \lstick{\mathrm{q_3}}&\qw      &\qw\\
  \lstick{\mathrm{q_4}}&\qw      &\qw\\&}
  \\
  \colrule
  Z1
  & 
  \Qcircuit @C=.5em @R=.5em @!R {
  \lstick{\mathrm{q_3}}&\gate{Z} &\qw\\
  \lstick{\mathrm{q_4}}&\qw      &\qw\\&}
  &
  \Qcircuit @C=.5em @R=.2em @!R {
  \lstick{\mathrm{q_1}}&\gate{Z} &\qw\\
  \lstick{\mathrm{q_2}}&\gate{Z} &\qw\\
  \lstick{\mathrm{q_3}}&\qw      &\qw\\
  \lstick{\mathrm{q_4}}&\qw      &\qw\\&}
  \\
  \colrule
  Z2
  & 
  \Qcircuit @C=.5em @R=.5em @!R {
  \lstick{\mathrm{q_3}}&\qw      &\qw\\
  \lstick{\mathrm{q_4}}&\gate{Z} &\qw\\&}
  &
  \Qcircuit @C=.5em @R=.2em @!R {
  \lstick{\mathrm{q_1}}&\gate{Z} &\qw\\
  \lstick{\mathrm{q_2}}&\qw      &\qw\\
  \lstick{\mathrm{q_3}}&\gate{Z} &\qw\\
  \lstick{\mathrm{q_4}}&\qw      &\qw\\&}
  \\
  \colrule
  HHS
  & 
  \Qcircuit @C=.5em @R=.5em @!R {
  \lstick{\mathrm{q_3}}&\gate{H} &\ctrl{1} &\gate{H} &\ctrl{1} &\gate{H} &\ctrl{1} &\qw\\
  \lstick{\mathrm{q_4}}&\gate{H} &\targ    &\gate{H} &\targ    &\gate{H} &\targ    &\qw\\&}
  &
  \Qcircuit @C=.5em @R=.2em @!R {
  \lstick{\mathrm{q_1}}&\gate{H} &\qw\\
  \lstick{\mathrm{q_2}}&\gate{H} &\qw\\
  \lstick{\mathrm{q_3}}&\gate{H} &\qw\\
  \lstick{\mathrm{q_4}}&\gate{H} &\qw\\&}
  \\
  \colrule
  CZ
  & 
  \Qcircuit @C=.5em @R=.5em @!R {
  \lstick{\mathrm{q_3}}&\qw      &\ctrl{1} &\qw      &\qw\\
  \lstick{\mathrm{q_4}}&\gate{H} &\targ    &\gate{H} &\qw\\&}
  &
  \Qcircuit @C=.5em @R=.2em @!R {
  \lstick{\mathrm{q_1}}&\gate{S} &\qw      &\qw\\
  \lstick{\mathrm{q_2}}&\gate{S} &\gate{Z} &\qw\\
  \lstick{\mathrm{q_3}}&\gate{S} &\gate{Z} &\qw\\
  \lstick{\mathrm{q_4}}&\gate{S} &\qw      &\qw\\&}
  \\
\end{tabular}
\end{ruledtabular}
\end{table}

\FloatBarrier
\clearpage

\section{Full set of tested circuits}\label{app:fullcircuits}

In \tabref{tab:fullcircuits}, we give a list of all 465 circuits generated by
the procedure suggested in \cite{Gottesman2016quantumfaulttolerance} for the
maximum circuit length $T=10$, the repetition parameter $\mathrm{RP}=6$, and the
periodicity $P=3$. 

\begin{figure*}
  \includegraphics[width=\linewidth]{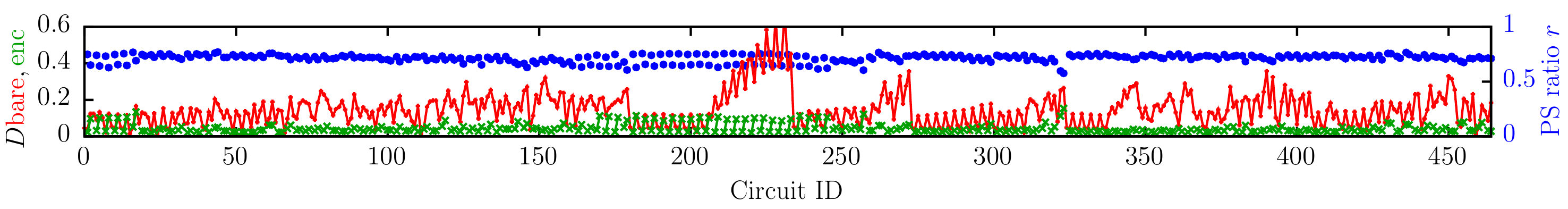}
  \caption{\label{ibmqx5full}Test of the fault-tolerance criterion in system (3)
  (see \secref{sec:ibmq}) for all 465 circuits using the qubits
  $(Q_4,Q_3,Q_2,Q_{15},Q_{14})$ of the IBM 16-qubit device \texttt{ibmqx5} on
  April 19, 2018. Shown are the statistical distances to the ideal result for
  the selected bare (red plusses) and encoded (green crosses) circuits, and the
  postselection ratios (blue dots).  
  All plotted quantities are dimensionless.
  Only the circuits that could be mapped on
  the topology were run on the real device. Lines connecting the data points are
guides to the eye.}
\end{figure*}

\begin{figure*}
  \includegraphics[width=\linewidth]{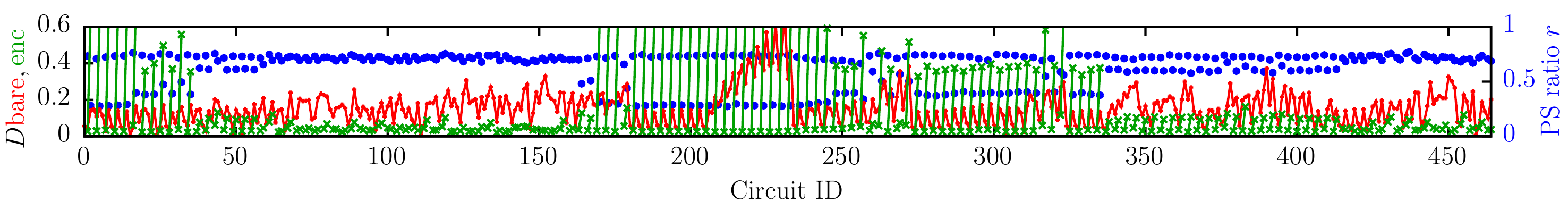}
  \caption{\label{ibmqx5full20}Test of the fault-tolerance criterion in system
  (3) (see \secref{sec:ibmq}) for all 465 circuits using the qubits
  $(Q_4,Q_3,Q_2,Q_{15},Q_{14})$ of the IBM 16-qubit device \texttt{ibmqx5} on
  April 20, 2018. Shown are the statistical distances to the ideal result for
  the selected bare (red plusses) and encoded (green crosses) circuits, and the
  postselection ratios (blue dots).  
  All plotted quantities are dimensionless.
  Only the circuits that could be mapped on
  the topology were run on the real device. Lines connecting the data points are
guides to the eye.}
\end{figure*}

A representative result of the performance of all circuits on
the IBM device is shown in \figref{ibmqx5full} (note that the
interruptions in the solid green line are due to the fact that the encoded
version of $\ket{00}$ cannot be prepared using the topology of system (3)).
This result undeniably demonstrates that encoding the circuits according to the
fault-tolerant scheme can improve the overall performance of the circuits that
can be implemented on the device. However, as already mentioned in
\secref{sec:ibmq}, the fault-tolerance criterion was not satisfied on all days
that we ran the experiment.  One such result is shown in \figref{ibmqx5full20}
where some of the encoded circuits with initial state $\ket{\Phi^+}$ have rather
high statistical distances and low PS ratios. 

\begin{figure*}
  \includegraphics[width=\linewidth]{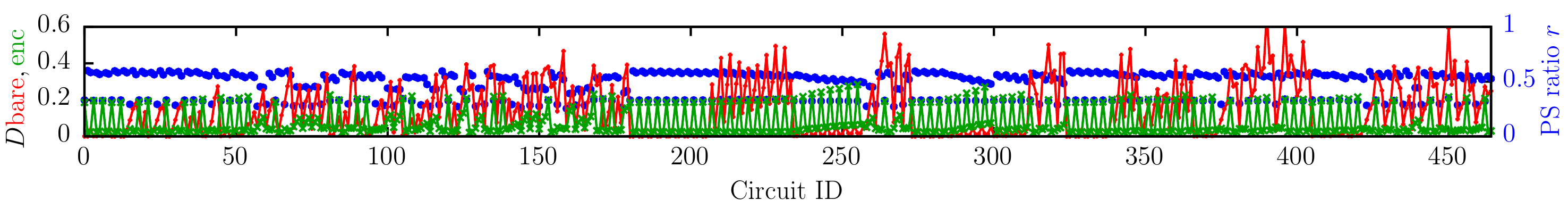}
  \caption{\label{envfull}Test of the fault-tolerance criterion in the
  decoherence model (system (1), see \secref{sec:spinqubits}) using $N_E=5$,
  $\beta=1$, and $\lambda=0.1$ for the full set of circuits. Shown are the
  statistical distances to the ideal result for the selected bare (red plusses)
  and encoded (green crosses) circuits, and the postselection ratios (blue
dots).
  All plotted quantities are dimensionless.
  Lines connecting the data points are guides to the eye.}
\end{figure*}

For completeness, we also present results for the full set of circuits tested in
the decoherence model (system (1)) in \figref{envfull}. This figure
does not have the above-mentioned interruptions since, in system (1), all 465
circuits can be implemented and tested in both their bare and encoded version.

\begin{table*}[h]
  \caption{\label{tab:fullcircuits} List of all 465 circuits used to test the
  fault-tolerance criterion. The elements consist of a range of three circuit
  IDs labeling the subsequent circuits, which consist of a particular set of
  gates operating on three initial states $\ket{i}\in\left(
  \ket{00},\ket{0+},\ket{\Phi^+} \right)$, enumerated in this order.}
\begin{ruledtabular}
\begin{tabular}{l}
\scriptsize
\begin{minipage}{0.3\linewidth}
\begin{verbatim}
0-2 |i>
3-5 X1 |i>
6-8 X2 |i>
9-11 Z1 |i>
12-14 Z2 |i>
15-17 HHS |i>
18-20 CZ |i>
21-23 X2 Z1 |i>
24-26 HHS Z1 |i>
27-29 Z1 Z2 |i>
30-32 X1 HHS |i>
33-35 CZ Z2 |i>
36-38 Z2 Z1 CZ |i>
39-41 Z1 X2 X2 |i>
42-44 CZ CZ HHS |i>
45-47 X1 X1 X1 |i>
48-50 Z2 X2 Z1 |i>
51-53 X1 X2 X1 |i>
54-56 X2 X1 CZ X1 |i>
57-59 HHS Z2 CZ Z1 |i>
60-62 HHS X1 Z2 Z2 |i>
63-65 CZ Z2 Z1 Z2 |i>
66-68 HHS HHS HHS Z1 |i>
69-71 X2 Z2 HHS CZ |i>
72-74 Z1 HHS CZ X2 Z2 |i>
75-77 X2 Z2 Z1 HHS CZ |i>
78-80 HHS X2 Z2 CZ CZ |i>
81-83 X1 X2 X1 X2 X1 |i>
84-86 Z2 Z1 X1 Z2 CZ |i>
87-89 HHS CZ HHS X2 CZ |i>
90-92 Z2 CZ X2 X2 X1 Z1 |i>
93-95 Z1 X2 Z1 X2 X1 Z1 |i>
96-98 Z1 Z2 X1 Z1 HHS X2 |i>
99-101 Z2 CZ X1 HHS X1 CZ |i>
102-104 X2 CZ HHS X2 CZ Z2 |i>
105-107 X1 X1 X1 X2 X2 Z2 |i>
108-110 Z1 X2 Z2 CZ X2 X1 X1 |i>
111-113 HHS X2 X2 Z2 Z2 X1 X1 |i>
114-116 Z2 X1 Z2 X2 CZ HHS CZ |i>
117-119 X2 Z2 Z1 HHS Z1 HHS HHS |i>
120-122 CZ Z2 Z1 Z2 X1 CZ X2 |i>
123-125 X2 HHS Z1 X1 X2 CZ X2 |i>
126-128 Z2 Z1 HHS HHS X2 X1 Z2 CZ |i>
129-131 Z1 X2 Z1 HHS CZ Z2 Z2 X2 |i>
132-134 CZ Z2 HHS Z2 HHS CZ Z2 HHS |i>
135-137 CZ X2 CZ CZ X2 X2 Z2 Z2 |i>
138-140 Z1 Z2 CZ CZ X1 X1 X2 X2 |i>
141-143 HHS X1 X2 X1 X2 Z2 Z1 X1 |i>
144-146 HHS CZ X2 HHS X1 X1 Z1 X1 X2 |i>
147-149 X1 X1 HHS Z2 HHS HHS X2 Z2 CZ |i>
150-152 X1 CZ HHS CZ HHS Z1 CZ CZ X2 |i>
153-155 CZ X1 Z2 HHS X2 X1 Z1 Z1 HHS |i>
\end{verbatim}
\end{minipage}
\begin{minipage}{0.345\linewidth}
\begin{verbatim}
156-158 X2 CZ HHS HHS HHS Z2 CZ CZ Z1 |i>
159-161 Z2 HHS CZ X2 X2 Z1 Z2 X1 X1 |i>
162-164 X2 X1 CZ HHS CZ Z1 Z1 X1 X2 Z2 |i>
165-167 Z2 X1 Z1 Z1 CZ Z1 X2 Z1 HHS CZ |i>
168-170 CZ HHS X1 Z2 X2 X2 X2 Z2 HHS CZ |i>
171-173 CZ X1 X2 Z1 Z1 X1 X1 Z1 Z1 Z2 |i>
174-176 Z2 Z2 Z2 X2 Z1 CZ CZ Z2 X2 X2 |i>
177-179 Z1 X2 HHS CZ X1 HHS CZ CZ X1 X1 |i>
180-182 Z2 Z2 |i>
183-185 Z2 Z2 Z2 |i>
186-188 Z2 Z2 Z2 Z2 |i>
189-191 Z2 Z2 Z2 Z2 Z2 |i>
192-194 Z2 Z2 Z2 Z2 Z2 Z2 |i>
195-197 Z2 Z2 Z2 Z2 Z2 Z2 Z2 |i>
198-200 Z2 Z2 Z2 Z2 Z2 Z2 Z2 Z2 |i>
201-203 Z2 Z2 Z2 Z2 Z2 Z2 Z2 Z2 Z2 |i>
204-206 Z2 Z2 Z2 Z2 Z2 Z2 Z2 Z2 Z2 Z2 |i>
207-209 CZ CZ |i>
210-212 CZ CZ CZ |i>
213-215 CZ CZ CZ CZ |i>
216-218 CZ CZ CZ CZ CZ |i>
219-221 CZ CZ CZ CZ CZ CZ |i>
222-224 CZ CZ CZ CZ CZ CZ CZ |i>
225-227 CZ CZ CZ CZ CZ CZ CZ CZ |i>
228-230 CZ CZ CZ CZ CZ CZ CZ CZ CZ |i>
231-233 CZ CZ CZ CZ CZ CZ CZ CZ CZ CZ |i>
234-236 X1 X1 |i>
237-239 X1 X1 X1 X1 |i>
240-242 X1 X1 X1 X1 X1 |i>
243-245 X1 X1 X1 X1 X1 X1 |i>
246-248 X1 X1 X1 X1 X1 X1 X1 |i>
249-251 X1 X1 X1 X1 X1 X1 X1 X1 |i>
252-254 X1 X1 X1 X1 X1 X1 X1 X1 X1 |i>
255-257 X1 X1 X1 X1 X1 X1 X1 X1 X1 X1 |i>
258-260 HHS CZ |i>
261-263 HHS CZ HHS CZ |i>
264-266 HHS CZ HHS CZ HHS CZ |i>
267-269 HHS CZ HHS CZ HHS CZ HHS CZ |i>
270-272 HHS CZ HHS CZ HHS CZ HHS CZ HHS CZ |i>
273-275 Z1 Z2 Z1 Z2 |i>
276-278 Z1 Z2 Z1 Z2 Z1 Z2 |i>
279-281 Z1 Z2 Z1 Z2 Z1 Z2 Z1 Z2 |i>
282-284 Z1 Z2 Z1 Z2 Z1 Z2 Z1 Z2 Z1 Z2 |i>
285-287 X2 X2 |i>
288-290 X2 X2 X2 X2 |i>
291-293 X2 X2 X2 X2 X2 X2 |i>
294-296 X2 X2 X2 X2 X2 X2 X2 X2 |i>
297-299 X2 X2 X2 X2 X2 X2 X2 X2 X2 X2 |i>
300-302 X2 Z1 X2 Z1 |i>
303-305 X2 Z1 X2 Z1 X2 Z1 |i>
306-308 X2 Z1 X2 Z1 X2 Z1 X2 Z1 |i>
309-311 X2 Z1 X2 Z1 X2 Z1 X2 Z1 X2 Z1 |i>
\end{verbatim}
\end{minipage}
\begin{minipage}{0.33\linewidth}
\begin{verbatim}
312-314 X1 HHS X1 HHS |i>
315-317 X1 HHS X1 HHS X1 HHS |i>
318-320 X1 HHS X1 HHS X1 HHS X1 HHS |i>
321-323 X1 HHS X1 HHS X1 HHS X1 HHS X1 HHS |i>
324-326 Z2 Z1 |i>
327-329 Z2 Z1 Z2 Z1 |i>
330-332 Z2 Z1 Z2 Z1 Z2 Z1 |i>
333-335 Z2 Z1 Z2 Z1 Z2 Z1 Z2 Z1 |i>
336-338 Z2 Z1 Z2 Z1 Z2 Z1 Z2 Z1 Z2 Z1 |i>
339-341 CZ CZ X1 |i>
342-344 CZ CZ X1 CZ CZ X1 |i>
345-347 CZ CZ X1 CZ CZ X1 CZ CZ X1 |i>
348-350 X1 CZ Z2 |i>
351-353 X1 CZ Z2 X1 CZ Z2 |i>
354-356 X1 CZ Z2 X1 CZ Z2 X1 CZ Z2 |i>
357-359 CZ CZ X2 |i>
360-362 CZ CZ X2 CZ CZ X2 |i>
363-365 CZ CZ X2 CZ CZ X2 CZ CZ X2 |i>
366-368 Z1 Z1 X1 |i>
369-371 Z1 Z1 X1 Z1 Z1 X1 |i>
372-374 Z1 Z1 X1 Z1 Z1 X1 Z1 Z1 X1 |i>
375-377 X2 Z2 HHS |i>
378-380 X2 Z2 HHS X2 Z2 HHS |i>
381-383 X2 Z2 HHS X2 Z2 HHS X2 Z2 HHS |i>
384-386 HHS HHS X2 |i>
387-389 HHS HHS X2 HHS HHS X2 |i>
390-392 HHS HHS X2 HHS HHS X2 HHS HHS X2 |i>
393-395 HHS CZ X2 CZ |i>
396-398 HHS CZ X2 CZ HHS CZ X2 CZ |i>
399-401 Z1 HHS HHS X2 |i>
402-404 Z1 HHS HHS X2 Z1 HHS HHS X2 |i>
405-407 Z2 Z1 X2 Z2 |i>
408-410 Z2 Z1 X2 Z2 Z2 Z1 X2 Z2 |i>
411-413 Z2 X1 Z2 X2 |i>
414-416 Z2 X1 Z2 X2 Z2 X1 Z2 X2 |i>
417-419 Z1 Z1 X2 X2 |i>
420-422 Z1 Z1 X2 X2 Z1 Z1 X2 X2 |i>
423-425 X2 HHS Z2 Z2 |i>
426-428 X2 HHS Z2 Z2 X2 HHS Z2 Z2 |i>
429-431 Z1 Z2 Z1 HHS X1 |i>
432-434 Z1 Z2 Z1 HHS X1 Z1 Z2 Z1 HHS X1 |i>
435-437 Z1 CZ Z1 Z2 HHS |i>
438-440 Z1 CZ Z1 Z2 HHS Z1 CZ Z1 Z2 HHS |i>
441-443 Z1 Z2 X2 Z1 HHS |i>
444-446 Z1 Z2 X2 Z1 HHS Z1 Z2 X2 Z1 HHS |i>
447-449 X1 Z2 HHS CZ CZ |i>
450-452 X1 Z2 HHS CZ CZ X1 Z2 HHS CZ CZ |i>
453-455 X1 X1 X1 HHS Z2 |i>
456-458 X1 X1 X1 HHS Z2 X1 X1 X1 HHS Z2 |i>
459-461 HHS X1 Z2 X1 Z2 |i>
462-464 HHS X1 Z2 X1 Z2 HHS X1 Z2 X1 Z2 |i>

\end{verbatim}
\end{minipage}
\normalsize
\end{tabular}
\end{ruledtabular}
\end{table*}
%\clearpage

\section{Specification of the simulation models}\label{app:simulationmodels}

\subsection{Spin qubits coupled to an environment}\label{app:spinqubits}

The model of five spin qubits coupled to an environment with $N_E$ two-level
defects is defined by the Hamiltonian given in \equsref{eq:H}{eq:HQE}. For this
Hamiltonian, we numerically solve the TDSE given in \equref{eq:TDSE} by means of
the Chebyshev polynomial algorithm to machine precision
\cite{talezerkosloff1984chebyshev, dobrovitski2003chebyshev,
deraedt2004computational}, which yields the state $\ket{\Psi(t)}$ of the system
after execution of a particular circuit. 

By construction, the only source of errors in this model is the interaction of
the qubits with the environment controlled by the coupling strength $\lambda$.
For $\lambda=0$, the quantum computer model is designed to work perfectly.
Therefore, the quantum gates in this model are not implemented by pulses but by
choosing suitable parameters $h_n^\alpha$ and $G_{nm}^\alpha$ for $H_Q$ given by
\equref{eq:HQ} \cite{deraedt2004computational}, and having the system evolve
through the TDSE given in \equref{eq:TDSE} for a certain time $t$. The specific
set of parameters for the gates used in the tested circuits is given in
\tabref{tab:envgateparameters}.  The two-qubit gate $\text{CNOT}_{nm}$ between
qubits $n$ and $m$ is implemented through the gate sequence $H_nI_{nm}H_n$,
where $H_n$ is the Hadamard gate on qubit $n$ and $I_{nm}$ implements a
two-qubit evolution of the form $\sigma_n^x+\sigma_m^x-\sigma_n^x\sigma_m^x$
through \equref{eq:HQ} (see \tabref{tab:envgateparameters}).

\begin{table}
  \caption{\label{tab:envgateparameters} Summary of the parameters for the
  required set of quantum gates (see \equsref{X1L}{CZL}), implemented through
  the time evolution of $H_Q$ given in \equref{eq:HQ}. Each parameter
  $h_n^\alpha$ and $G_{nm}^\alpha$ is given in GHz, and the duration $t$ of the
corresponding gate is given in ns.}
\begin{ruledtabular}
\begin{tabular}{@{}cccccc@{}}
  Gate & $h_n^x$ & $h_m^x$ & $h_n^z$ & $G_{nm}^x$ & $t$ \\
  \colrule
   $X_n$ & 1 & - & 0 & 0 & $\pi/2$ \\
   $Z_n$ & 0 & - & $15+n/2$ & 0 & $\pi/(30+n)$ \\
   $S_n$ & 0 & - & $15+n/2$ & 0 & $\pi/(60+2n)$ \\
   $H_n$ & $15+n/2$ & - & $15+n/2$ & 0 & $\pi/\sqrt{2}/(30+n)$ \\
   $I_{nm}$ & $-0.025$ & $-0.025$ & 0 & 0.025 & $10\pi$ \\
\end{tabular}
\end{ruledtabular}
\end{table}

\begin{figure}
  \includegraphics[width=.5\linewidth]{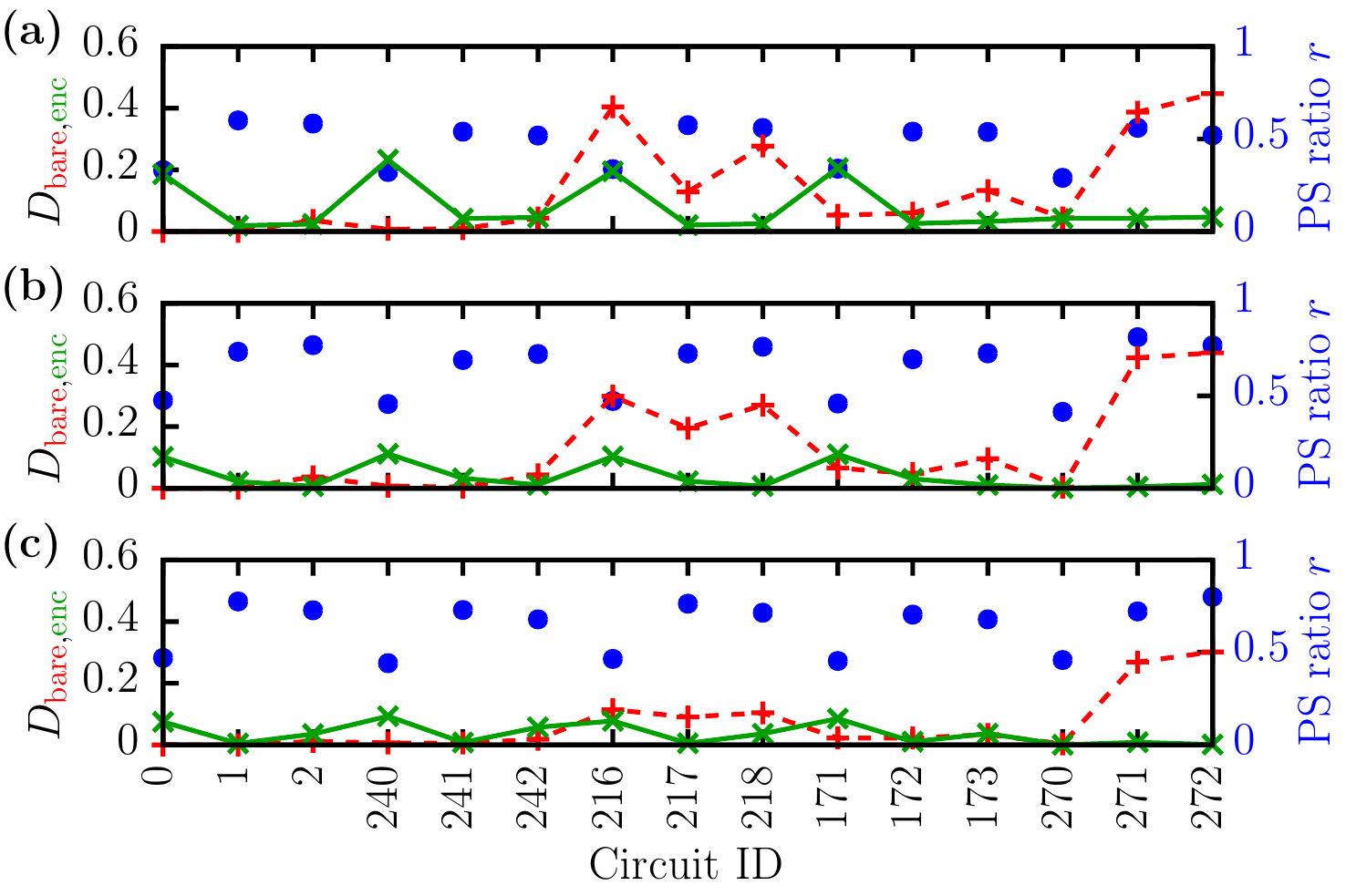}
  \caption{\label{circuitsNE} (Color online) Test of the fault-tolerance
  criterion for different environment sizes (a) $N_E=5$, (b) $N_E=20$, and (c)
  $N_E=27$. Shown are the statistical distances to the ideal result for the
  selected bare (dashed red line) and encoded (solid green line) circuits, and
  the postselection ratios (blue dots). 
  All plotted quantities are dimensionless.
  The simulations were done for inverse
  temperature $\beta=1$ and coupling strength $\lambda=0.1$. Lines connecting
  the data points are guides to the eye.}
\end{figure}

\begin{figure}
  \includegraphics[width=.5\linewidth]{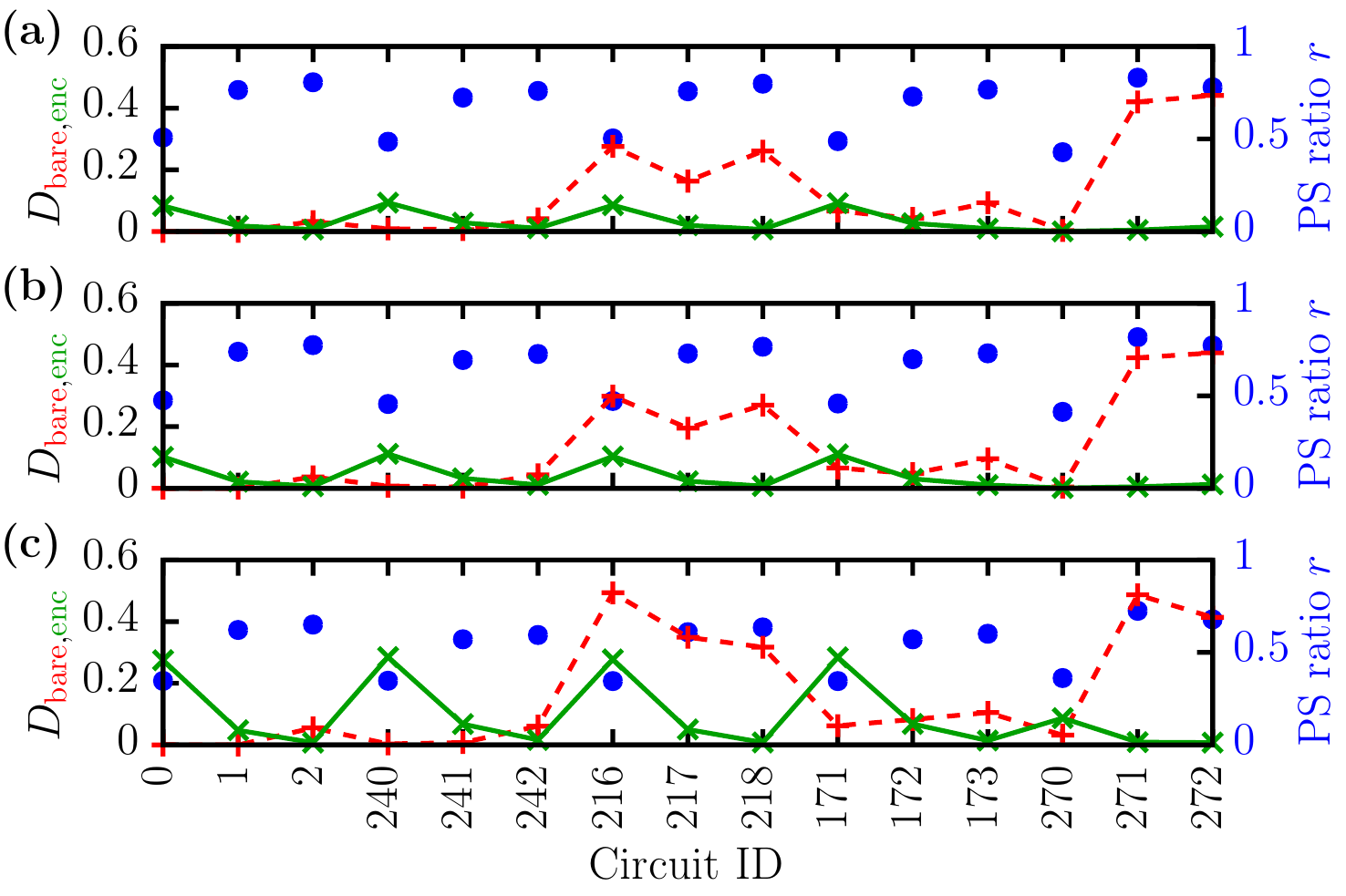}
  \caption{\label{circuitsbeta} (Color online) Test of the fault-tolerance
  criterion for different inverse temperatures (a) $\beta=0$, (b) $\beta=1$, and (c)
  $\beta=5$.  Shown are the statistical distances to the ideal result for the
  selected bare (dashed red line) and encoded (solid green line) circuits, and
the postselection ratios (blue dots).  
  All plotted quantities are dimensionless.
All simulations were done for
environment size $N_E=20$ and coupling strength $\lambda=0.1$. Lines connecting
the data points are guides to the eye.}
\end{figure}

In addition to the results for $\lambda\in\left[ 0.01,0.2 \right]$ presented in
\secref{sec:spinqubits}, we have studied the performance of the circuits for
various numbers $N_E$ of two-level systems in the environment (see
\figref{circuitsNE}) and various inverse temperatures $\beta$ (see
\figref{circuitsbeta}). 

The number $N_E$ of two-level systems in the environment is
limited by two factors. On the one hand, it should not be too small for
decoherence effects to be observable, so $N_E=5$ is the smallest case that we
consider. In this case, the virtual interaction between two spin qubits mediated
by the environment has a significantly smaller path than for larger $N_E$, so
the influence of the environment on the performance is rather strong (see
\figref{circuitsNE}(a)). On the other hand, there is a practical limitation
given by the available computational resources on the supercomputer.  For
$N_E=27$, the dimension of the total Hilbert space is $2^{32}$, so the
simulation of the full time evolution is rather expensive. Hence this is the
bound we set for what can be simulated with a reasonable amount of computer
resources (CPU time and memory). By comparing the results for $N_E=20$ to those
for $N_E=27$, shown in \figref{circuitsNE}(b) and (c) respectively, we find that
there is no significant qualitative change. In particular, the statistical
distances $D_{\text{enc}}$ for the encoded circuits all lie between $0$ and
$0.15$, with only small fluctuations between the results for $N_E=20$ and
$N_E=27$. But the main observation is that in all cases, some circuits perform
better when they are encoded while others are better without encoding.

The dependence of the fault-tolerance test on the inverse temperature $\beta$ is
shown in \figref{circuitsbeta}. We find no significant difference
between the results for $\beta=0$ and $\beta=1$ shown in
\figref{circuitsbeta}(a) and (b), respectively. These cases represent the high
temperature regime. In contrast, the results for $\beta=5$, shown in
\figref{circuitsbeta}(c), resemble the results from the smallest environment
$N_E=5$, shown in \figref{circuitsNE}(a). Hence, this low-temperature regime
indicates that the system is no longer affected by pure decoherence, but that
other effects also come into play. However, although some influence of the
temperature on the performance of the circuits can be observed, the qualitative
results do not change. This means that the bare circuits that outperform their
encoded equivalents are the same in each case. Hence, in this system, the
criterion for fault tolerance is not satisfied in any of the regimes under
investigation.

\subsection{Transmon simulation}\label{app:transmons}

The Hamiltonian given in \equsref{eq:modelhamiltonian}{eq:modelhamiltonianres}
models five transmon qubits coupled by six resonators, as schematically shown in
\figref{topology}. The full set of relevant device parameters is summarized in
\tabref{tab:transmonparemeters} and \tabref{tab:resonatorparemeters}. We
numerically solve the TDSE in \equref{eq:TDSE} for the time-dependent
Hamiltonian given in \equref{eq:modelhamiltonian}, using the unconditionally
stable Suzuki-Trotter product-formula algorithm \cite{deraedt1987productformula,
deraedt2004computational} to second order, to obtain $\ket{\Psi(t)}$ at any time
$t$. In this work, we express the full state $\ket{\Psi(t)}$ in the joint
eigenbasis of the five transmons and the six resonators $\ket{m_0\cdots
m_4;k_0\cdots k_5}$, where each transmon index $m_i$ enumerates the first four
eigenstates of $4E_{Ci}\hat{n}_i^2-E_{Ji}\cos\hat{\varphi}_i$, and the resonator
indices $k_r$ enumerate the first four Fock states. The algorithm results in
four-component updates of the full state $\ket{\Psi(t)}$ at each time step
$\tau=\SI{0.001}{ns}$. We have verified that this basis accurately covers the
system dynamics by comparison with exact diagonalization and with the simulation
in the charge basis (see \cite{Willsch2017GateErrorAnalysis} where 17 levels
were included for each transmon).

Quantum gates are implemented by choosing a particular pulse for $n_{gi}(t)$ in
the time-dependent Hamiltonian given in \equref{eq:modelhamiltoniantr}. As
in the corresponding experiments \cite{gambetta2013controlIFF, McKay2016VZgate},
a single-qubit pulse on qubit $i$ is defined by
\begin{align}
  \label{eq:singlequbitpulse}
  n_{gi}(t) &= \Omega_{\text{G}}(t)\cos(2\pi ft-\gamma)
  + \beta_X\dot\Omega_{\text{G}}(t)\cos(2\pi ft-\gamma-\frac{\pi}{2}),
\end{align}
where $\Omega_{\text{G}}(t)$ is a Gaussian with amplitude $\Omega_X$, duration
$T_X=\SI{80}{ns}$, and width $\sigma=T_X/4$ (see
\cite{Willsch2017GateErrorAnalysis}), $\beta_X$ is the DRAG coefficient
\cite{motzoi2009drag, gambetta2010dragtheory}, $f$ is the drive frequency, and
$\gamma$ is a phase parameter used to implement VZ gates \cite{McKay2016VZgate}.
The two-qubit CNOT gate is implemented using an echoed cross-resonance scheme
\cite{gambetta2013controlIFF, Willsch2017GateErrorAnalysis,
sheldon2016procedure}, in which the single-qubit pulses implementing the echo
are realized by \equref{eq:singlequbitpulse}, and the flat-top Gaussians are
obtained from the same equation by choosing $\beta_X=0$ and $\Omega_G(t)$ to
rise for $0\le t\le\SI{15}{ns}$ with $\sigma=\SI{5}{ns}$, stay constant at
$\Omega_{CR}$ for $\SI{15}{ns}<t<\SI{15}{ns}+T_{CR}$, and fall again for
$\SI{15}{ns}+T_{CR}\le t\le \SI{30}{ns}+T_{CR}$ (see
\cite{Willsch2017GateErrorAnalysis} for more information). 

We optimize two sets of gate pulses for the experiments, namely one
without frequency tuning and one with frequency tuning.  The pulses without
frequency tuning use the qubit frequencies as drive frequencies, i.e.,
$f=\omega/2\pi$, where the qubit frequencies $\omega$ are taken from
\tabref{tab:transmonparemeters}. In contrast, with frequency tuning, the pulse-optimization 
procedure also fits the drive frequencies, such that the resulting
pulses have the drive frequencies $f=\omega^{\text{dr}}/2\pi$ (the corresponding
values are also listed in \tabref{tab:transmonparemeters}). To distinguish
between both gate sets, we attach the suffix \texttt{-withf} to the pulses with
frequency tuning. The relevant parameters resulting from a Nelder-Mead
optimization \cite{NelderMead1965} are summarized in
\tabref{tab:xgateparameters} and \tabref{tab:cnotgateparameters}.

For each pulse, we evaluate various gate metrics such as the matrix distance
$\Delta$ used as the objective function in the optimization
\cite{Willsch2017GateErrorAnalysis}, the diamond distance $\eta_\Diamond$
\cite{kitaev1997diamondnorm}, the average gate fidelity $F_{\text{avg}}$
\cite{nielsen2002gatefidelity}, and the unitarity $u$
\cite{Wallman2015unitarity}. These metrics are reported in
\tabref{tab:gatemetrics}.

\begin{table}
  \caption{\label{tab:transmonparemeters} Device parameters of the transmon
  Hamiltonian defined in \equref{eq:modelhamiltoniantr}. All values are given in
  GHz. The charging energies $E_C$ and the Josephson energies $E_J$ define the
  transmon qubits. The qubit frequencies $\omega$ have been obtained by
  preparing the respective qubit in the state $\ket{+}$ and all other qubits in
  the state $\ket{0}$, having the entire system evolve for $\SI{1000}{ns}$, and
  measuring the frequency of $\expect{\sigma^x(t)}$. The drive frequencies
  $\omega^{\text{dr}}$ result from additionally tuning this $\omega$ in the
  single-qubit pulse-optimization procedure. These frequencies are
  only used by the gate set labeled \texttt{*-withf}.}
\begin{ruledtabular}
\begin{tabular}{@{}cccccc@{}}
  & $q_0$ & $q_1$ & $q_2$ & $q_3$ & $q_4$ \\
  \colrule
  $E_C/2\pi$ & 0.301 & 0.301 & 0.301 & 0.301 & 0.301 \\
  $E_J/2\pi$ & 11.6671 & 12.1273 & 13.003 & 12.2456 & 11.1943 \\
  $\omega/2\pi$ & 4.97154 & 5.07063 & 5.26657 & 5.10145 & 4.86036 \\
  $\omega^{\text{dr}}/2\pi$ & 4.97164 & 5.07043 & 5.26634 & 5.10147 & 4.86055
\end{tabular}
\end{ruledtabular}
\end{table}

\begin{table}
  \caption{\label{tab:resonatorparemeters} Device parameters of the resonator
  Hamiltonian defined in \equref{eq:modelhamiltonianres}. All values are given
in GHz.}
\begin{ruledtabular}
\begin{tabular}{@{}ccccccc@{}}
  & $r_0$ & $r_1$ & $r_2$ & $r_3$ & $r_4$ & $r_5$ \\
  \colrule
  $\Omega/2\pi$ & 6.45 & 6.25 & 6.65 & 6.65 & 6.45 & 6.85 \\
  $G/2\pi$ & 0.07 & 0.07 & 0.07 & 0.07 & 0.07 & 0.07 \\
  Coupled to & $q_1,q_2$ & $q_0,q_1$ & $q_2,q_3$ & $q_1,q_4$ & $q_3,q_4$ &
  $q_0,q_4$
\end{tabular}
\end{ruledtabular}
\end{table}

\begin{table}
  \caption{\label{tab:xgateparameters} Parameters of the Gaussian
  \texttt{xpih-*} pulses defined by \equref{eq:singlequbitpulse}. The drive
  frequencies $f$ are given in GHz and the pulse time $T_X$ and the DRAG
  coefficient $\beta_X$ are given in ns. The Gaussian drive amplitudes
  $\Omega_X$ are unitless. All pulses labeled \texttt{*-withf} represent pulses
  with frequency tuning, meaning that the drive frequency has additionally been
  optimized in the pulse-optimization
procedure.}
\begin{ruledtabular}
\begin{tabular}{@{}ccccc@{}}
  Pulse name & $f$ & $T_X$ & $\Omega_X$ & $\beta_X$ \\
  \colrule
  \texttt{xpih-0} & 4.97154 & 80 & 0.00238 & 1.335 \\
  \texttt{xpih-1} & 5.07063 & 80 & 0.00236 & -1.904 \\
  \texttt{xpih-2} & 5.26657 & 80 & 0.00233 & -2.165 \\
  \texttt{xpih-3} & 5.10145 & 80 & 0.00236 & 0.498 \\
  \texttt{xpih-4} & 4.86036 & 80 & 0.00241 & 2.276 \\
  \colrule
  \texttt{xpih-0-withf} & 4.97164 & 80 & 0.00239 & 0.239 \\
  \texttt{xpih-1-withf} & 5.07043 & 80 & 0.00236 & 0.238 \\
  \texttt{xpih-2-withf} & 5.26634 & 80 & 0.00233 & 0.229 \\
  \texttt{xpih-3-withf} & 5.10147 & 80 & 0.00236 & 0.232 \\
  \texttt{xpih-4-withf} & 4.86055 & 80 & 0.00241 & 0.236 \\
\end{tabular}
\end{ruledtabular}
\end{table}

\begin{table*} \caption{\label{tab:cnotgateparameters} Parameters defining the
  echoed cross-resonance pulses (CR2 in \cite{Willsch2017GateErrorAnalysis}) to
  implement the CNOT gate. The drive frequencies $f_C$ and $f_T$ are given in
  GHz. The times $T_{CR}$ of the flat top in a cross-resonance pulse, the
  Gaussian pulse times $T_X$, and the DRAG coefficients $\beta_C$ and $\beta_T$
  are given in ns. The Gaussian drive amplitudes $\Omega_{CR}$ and $\Omega_C$ on
  the control qubit and $\Omega_T$ on the target qubit are unitless. All pulses
  labeled \texttt{*-withf} represent pulses with frequency tuning, meaning that
the drive frequency has also been optimized in the pulse-optimization
procedure.}
\begin{ruledtabular}
\begin{tabular}{@{}cccccccccc@{}}
  Pulse name & $f_C$ & $f_T$ & $T_{CR}$ & $T_X$ & $\Omega_{CR}$ & $\Omega_C$ & $\beta_C$ & $\Omega_T$ & $\beta_T$ \\
  \colrule
  \texttt{cnot-1-0} & 5.07063 & 4.97154 & 76.955 & 80 & 0.0097 & 0.00461 & 0.640 & 0.00238 & 1.335 \\
  \texttt{cnot-1-4} & 5.07063 & 4.86036 & 64.161 & 80 & 0.0183 & 0.00476 & -0.148 & 0.00241 & 2.276 \\
  \texttt{cnot-2-1} & 5.26657 & 5.07063 & 33.398 & 80 & 0.0235 & 0.00465 & -0.036 & 0.00236 & -1.904 \\
  \texttt{cnot-3-2} & 5.10145 & 5.26657 & 242.064 & 80 & 0.0111 & 0.00471 & 0.508 & 0.00233 & -2.165 \\
  \texttt{cnot-3-4} & 5.10145 & 4.86036 & 33.247 & 80 & 0.0290 & 0.00465 & 0.640 & 0.00241 & 2.276 \\
  \texttt{cnot-4-0} & 4.86036 & 4.97154 & 105.151 & 80 & 0.0210 & 0.00449 & -1.511 & 0.00238 & 1.335 \\
  \colrule
  \texttt{cnot-1-0-withf} & 5.07043 & 4.97164 & 73.538 & 80 & 0.0101 & 0.00477 & 0.798 & 0.00239 & 0.239 \\
  \texttt{cnot-1-4-withf} & 5.07043 & 4.86055 & 109.439 & 80 & 0.0114 & 0.00472 & 0.502 & 0.00241 & 0.236 \\
  \texttt{cnot-2-1-withf} & 5.26634 & 5.07043 & 82.077 & 80 & 0.0111 & 0.00463 & 0.661 & 0.00236 & 0.238 \\
  \texttt{cnot-3-2-withf} & 5.10147 & 5.26634 & 58.763 & 80 & 0.0429 & 0.00480 & -0.198 & 0.00233 & 0.229 \\
  \texttt{cnot-3-4-withf} & 5.10147 & 4.86055 & 85.294 & 80 & 0.0118 & 0.00474 & 0.247 & 0.00241 & 0.236 \\
  \texttt{cnot-4-0-withf} & 4.86055 & 4.97164 & 98.599 & 80 & 0.0239 & 0.00483 & 0.115 & 0.00239 & 0.239 \\
\end{tabular}
\end{ruledtabular}
\end{table*}

\begin{table}
  \caption{\label{tab:gatemetrics} Gate metrics resulting from the pulse-optimization 
  procedure. $\Delta$ is the distance objective (loss function),
  $\eta_\Diamond$ is the diamond distance, $F_{\text{avg}}$ is the average gate
  fidelity, and $u$ is the unitarity (see \cite{Willsch2017GateErrorAnalysis}
for more information about these metrics).}
\begin{ruledtabular}
\begin{tabular}{@{}ccccc@{}}
  Pulse name & $\Delta$ & $\eta_\Diamond$ & $F_{\text{avg}}$ & $u$ \\
  \colrule
  \texttt{xpih-0} & $4.60\e{-5}$ & 0.007 & 0.9930 & 0.9860 \\
  \texttt{xpih-1} & $1.19\e{-4}$ & 0.011 & 0.9884 & 0.9770 \\
  \texttt{xpih-2} & $7.52\e{-6}$ & 0.002 & 0.9962 & 0.9925 \\
  \texttt{xpih-3} & $8.99\e{-6}$ & 0.003 & 0.9965 & 0.9930 \\
  \texttt{xpih-4} & $4.17\e{-5}$ & 0.006 & 0.9934 & 0.9868 \\
  \colrule
  \texttt{xpih-0-withf} & $4.59\e{-5}$ & 0.007 & 0.9930 & 0.9860 \\
  \texttt{xpih-1-withf} & $1.14\e{-4}$ & 0.011 & 0.9887 & 0.9774 \\
  \texttt{xpih-2-withf} & $7.20\e{-6}$ & 0.002 & 0.9963 & 0.9927 \\
  \texttt{xpih-3-withf} & $8.85\e{-6}$ & 0.003 & 0.9965 & 0.9930 \\
  \texttt{xpih-4-withf} & $3.87\e{-5}$ & 0.006 & 0.9936 & 0.9873 \\
  \colrule
  \texttt{cnot-1-0} & $1.34\e{-2}$ & 0.071 & 0.9852 & 0.9758 \\
  \texttt{cnot-1-4} & $1.08\e{-1}$ & 0.177 & 0.9621 & 0.9668 \\
  \texttt{cnot-2-1} & $4.68\e{-2}$ & 0.119 & 0.9714 & 0.9615 \\
  \texttt{cnot-3-2} & $1.83\e{-2}$ & 0.088 & 0.9852 & 0.9777 \\
  \texttt{cnot-3-4} & $9.54\e{-2}$ & 0.179 & 0.9671 & 0.9720 \\
  \texttt{cnot-4-0} & $2.78\e{-1}$ & 0.284 & 0.9347 & 0.9783 \\
  \colrule
  \texttt{cnot-1-0-withf} & $5.70\e{-2}$ & 0.149 & 0.9751 & 0.9728 \\
  \texttt{cnot-1-4-withf} & $7.13\e{-3}$ & 0.056 & 0.9841 & 0.9712 \\
  \texttt{cnot-2-1-withf} & $1.38\e{-2}$ & 0.081 & 0.9806 & 0.9668 \\
  \texttt{cnot-3-2-withf} & $1.21\e{-1}$ & 0.207 & 0.9644 & 0.9764 \\
  \texttt{cnot-3-4-withf} & $1.88\e{-2}$ & 0.090 & 0.9832 & 0.9740 \\
  \texttt{cnot-4-0-withf} & $8.27\e{-2}$ & 0.168 & 0.9739 & 0.9806 \\
\end{tabular}
\end{ruledtabular}
\end{table}

\FloatBarrier
\twocolumngrid

\bibliographystyle{apsrev4-1custom}
\bibliography{bibliography}

\begin{thebibliography}{60}%
\makeatletter
\providecommand \@ifxundefined [1]{%
 \@ifx{#1\undefined}
}%
\providecommand \@ifnum [1]{%
 \ifnum #1\expandafter \@firstoftwo
 \else \expandafter \@secondoftwo
 \fi
}%
\providecommand \@ifx [1]{%
 \ifx #1\expandafter \@firstoftwo
 \else \expandafter \@secondoftwo
 \fi
}%
\providecommand \natexlab [1]{#1}%
\providecommand \enquote  [1]{``#1''}%
\providecommand \bibnamefont  [1]{#1}%
\providecommand \bibfnamefont [1]{#1}%
\providecommand \citenamefont [1]{#1}%
\providecommand \href@noop [0]{\@secondoftwo}%
\providecommand \href [0]{\begingroup \@sanitize@url \@href}%
\providecommand \@href[1]{\@@startlink{#1}\@@href}%
\providecommand \@@href[1]{\endgroup#1\@@endlink}%
\providecommand \@sanitize@url [0]{\catcode `\\12\catcode `\$12\catcode
  `\&12\catcode `\#12\catcode `\^12\catcode `\_12\catcode `\%12\relax}%
\providecommand \@@startlink[1]{}%
\providecommand \@@endlink[0]{}%
\providecommand \url  [0]{\begingroup\@sanitize@url \@url }%
\providecommand \@url [1]{\endgroup\@href {#1}{\urlprefix }}%
\providecommand \urlprefix  [0]{URL }%
\providecommand \Eprint [0]{\href }%
\providecommand \doibase [0]{http://dx.doi.org/}%
\providecommand \selectlanguage [0]{\@gobble}%
\providecommand \bibinfo  [0]{\@secondoftwo}%
\providecommand \bibfield  [0]{\@secondoftwo}%
\providecommand \translation [1]{[#1]}%
\providecommand \BibitemOpen [0]{}%
\providecommand \bibitemStop [0]{}%
\providecommand \bibitemNoStop [0]{.\EOS\space}%
\providecommand \EOS [0]{\spacefactor3000\relax}%
\providecommand \BibitemShut  [1]{\csname bibitem#1\endcsname}%
\let\auto@bib@innerbib\@empty
%</preamble>
\bibitem [{\citenamefont {Sheldon}\ \emph
  {et~al.}(2016{\natexlab{a}})\citenamefont {Sheldon}, \citenamefont {Bishop},
  \citenamefont {Magesan}, \citenamefont {Filipp}, \citenamefont {Chow},\ and\
  \citenamefont {Gambetta}}]{sheldon2015singlequbitfidelities}%
  \BibitemOpen
  \bibfield  {author} {\bibinfo {author} {\bibfnamefont {S.}~\bibnamefont
  {Sheldon}}, \bibinfo {author} {\bibfnamefont {L.~S.}\ \bibnamefont {Bishop}},
  \bibinfo {author} {\bibfnamefont {E.}~\bibnamefont {Magesan}}, \bibinfo
  {author} {\bibfnamefont {S.}~\bibnamefont {Filipp}}, \bibinfo {author}
  {\bibfnamefont {J.~M.}\ \bibnamefont {Chow}}, \ and\ \bibinfo {author}
  {\bibfnamefont {J.~M.}\ \bibnamefont {Gambetta}},\ }\bibfield  {title}
  {\enquote {\bibinfo {title} {Characterizing errors on qubit operations via
  iterative randomized benchmarking},}\ }\href@noop {} {\bibfield  {journal}
  {\bibinfo  {journal} {Phys. Rev. A}\ }\textbf {\bibinfo {volume} {93}},\
  \bibinfo {pages} {012301} (\bibinfo {year} {2016}{\natexlab{a}})}\BibitemShut
  {NoStop}%
\bibitem [{\citenamefont {Gambetta}\ \emph {et~al.}(2017)\citenamefont
  {Gambetta}, \citenamefont {Chow},\ and\ \citenamefont
  {Steffen}}]{gambetta2015building}%
  \BibitemOpen
  \bibfield  {author} {\bibinfo {author} {\bibfnamefont {J.~M.}\ \bibnamefont
  {Gambetta}}, \bibinfo {author} {\bibfnamefont {J.~M.}\ \bibnamefont {Chow}},
  \ and\ \bibinfo {author} {\bibfnamefont {M.}~\bibnamefont {Steffen}},\
  }\bibfield  {title} {\enquote {\bibinfo {title} {Building logical qubits in a
  superconducting quantum computing system},}\ }\href {\doibase
  10.1038/s41534-016-0004-0} {\bibfield  {journal} {\bibinfo  {journal} {npj
  Quantum Inf.}\ }\textbf {\bibinfo {volume} {3}},\ \bibinfo {pages} {2}
  (\bibinfo {year} {2017})}\BibitemShut {NoStop}%
\bibitem [{\citenamefont {Neill}\ \emph {et~al.}(2018)\citenamefont {Neill},
  \citenamefont {Roushan}, \citenamefont {Kechedzhi}, \citenamefont {Boixo},
  \citenamefont {Isakov}, \citenamefont {Smelyanskiy}, \citenamefont {Megrant},
  \citenamefont {Chiaro}, \citenamefont {Dunsworth}, \citenamefont {Arya},
  \citenamefont {Barends}, \citenamefont {Burkett}, \citenamefont {Chen},
  \citenamefont {Chen}, \citenamefont {Fowler}, \citenamefont {Foxen},
  \citenamefont {Giustina}, \citenamefont {Graff}, \citenamefont {Jeffrey},
  \citenamefont {Huang}, \citenamefont {Kelly}, \citenamefont {Klimov},
  \citenamefont {Lucero}, \citenamefont {Mutus}, \citenamefont {Neeley},
  \citenamefont {Quintana}, \citenamefont {Sank}, \citenamefont {Vainsencher},
  \citenamefont {Wenner}, \citenamefont {White}, \citenamefont {Neven},\ and\
  \citenamefont {Martinis}}]{Neil2017GoogleBlueprintQuantumSupremacy}%
  \BibitemOpen
  \bibfield  {author} {\bibinfo {author} {\bibfnamefont {C.}~\bibnamefont
  {Neill}}, \bibinfo {author} {\bibfnamefont {P.}~\bibnamefont {Roushan}},
  \bibinfo {author} {\bibfnamefont {K.}~\bibnamefont {Kechedzhi}}, \bibinfo
  {author} {\bibfnamefont {S.}~\bibnamefont {Boixo}}, \bibinfo {author}
  {\bibfnamefont {S.~V.}\ \bibnamefont {Isakov}}, \bibinfo {author}
  {\bibfnamefont {V.}~\bibnamefont {Smelyanskiy}}, \bibinfo {author}
  {\bibfnamefont {A.}~\bibnamefont {Megrant}}, \bibinfo {author} {\bibfnamefont
  {B.}~\bibnamefont {Chiaro}}, \bibinfo {author} {\bibfnamefont
  {A.}~\bibnamefont {Dunsworth}}, \bibinfo {author} {\bibfnamefont
  {K.}~\bibnamefont {Arya}}, \bibinfo {author} {\bibfnamefont {R.}~\bibnamefont
  {Barends}}, \bibinfo {author} {\bibfnamefont {B.}~\bibnamefont {Burkett}},
  \bibinfo {author} {\bibfnamefont {Y.}~\bibnamefont {Chen}}, \bibinfo {author}
  {\bibfnamefont {Z.}~\bibnamefont {Chen}}, \bibinfo {author} {\bibfnamefont
  {A.}~\bibnamefont {Fowler}}, \bibinfo {author} {\bibfnamefont
  {B.}~\bibnamefont {Foxen}}, \bibinfo {author} {\bibfnamefont
  {M.}~\bibnamefont {Giustina}}, \bibinfo {author} {\bibfnamefont
  {R.}~\bibnamefont {Graff}}, \bibinfo {author} {\bibfnamefont
  {E.}~\bibnamefont {Jeffrey}}, \bibinfo {author} {\bibfnamefont
  {T.}~\bibnamefont {Huang}}, \bibinfo {author} {\bibfnamefont
  {J.}~\bibnamefont {Kelly}}, \bibinfo {author} {\bibfnamefont
  {P.}~\bibnamefont {Klimov}}, \bibinfo {author} {\bibfnamefont
  {E.}~\bibnamefont {Lucero}}, \bibinfo {author} {\bibfnamefont
  {J.}~\bibnamefont {Mutus}}, \bibinfo {author} {\bibfnamefont
  {M.}~\bibnamefont {Neeley}}, \bibinfo {author} {\bibfnamefont
  {C.}~\bibnamefont {Quintana}}, \bibinfo {author} {\bibfnamefont
  {D.}~\bibnamefont {Sank}}, \bibinfo {author} {\bibfnamefont {A.}~\bibnamefont
  {Vainsencher}}, \bibinfo {author} {\bibfnamefont {J.}~\bibnamefont {Wenner}},
  \bibinfo {author} {\bibfnamefont {T.~C.}\ \bibnamefont {White}}, \bibinfo
  {author} {\bibfnamefont {H.}~\bibnamefont {Neven}}, \ and\ \bibinfo {author}
  {\bibfnamefont {J.~M.}\ \bibnamefont {Martinis}},\ }\bibfield  {title}
  {\enquote {\bibinfo {title} {A blueprint for demonstrating quantum supremacy
  with superconducting qubits},}\ }\href {\doibase 10.1126/science.aao4309}
  {\bibfield  {journal} {\bibinfo  {journal} {Science}\ }\textbf {\bibinfo
  {volume} {360}},\ \bibinfo {pages} {195} (\bibinfo {year}
  {2018})}\BibitemShut {NoStop}%
\bibitem [{\citenamefont {IBM}(2016)}]{ibmquantumexperience2016}%
  \BibitemOpen
  \bibfield  {author} {\bibinfo {author} {\bibnamefont {IBM}},\ }\href
  {https://www.research.ibm.com/ibm-q/} {\enquote {\bibinfo {title} {Q
  experience},}\ }\bibinfo {howpublished}
  {\url{https://www.research.ibm.com/ibm-q/}} (\bibinfo {year}
  {2016})\BibitemShut {NoStop}%
\bibitem [{\citenamefont {Willsch}\ \emph {et~al.}(2017)\citenamefont
  {Willsch}, \citenamefont {Nocon}, \citenamefont {Jin}, \citenamefont {{De
  Raedt}},\ and\ \citenamefont {Michielsen}}]{Willsch2017GateErrorAnalysis}%
  \BibitemOpen
  \bibfield  {author} {\bibinfo {author} {\bibfnamefont {D.}~\bibnamefont
  {Willsch}}, \bibinfo {author} {\bibfnamefont {M.}~\bibnamefont {Nocon}},
  \bibinfo {author} {\bibfnamefont {F.}~\bibnamefont {Jin}}, \bibinfo {author}
  {\bibfnamefont {H.}~\bibnamefont {{De Raedt}}}, \ and\ \bibinfo {author}
  {\bibfnamefont {K.}~\bibnamefont {Michielsen}},\ }\bibfield  {title}
  {\enquote {\bibinfo {title} {Gate-error analysis in simulations of quantum
  computers with transmon qubits},}\ }\href {\doibase
  10.1103/PhysRevA.96.062302} {\bibfield  {journal} {\bibinfo  {journal} {Phys.
  Rev. A}\ }\textbf {\bibinfo {volume} {96}},\ \bibinfo {pages} {062302}
  (\bibinfo {year} {2017})}\BibitemShut {NoStop}%
\bibitem [{\citenamefont {Michielsen}\ \emph {et~al.}(2017)\citenamefont
  {Michielsen}, \citenamefont {Nocon}, \citenamefont {Willsch}, \citenamefont
  {Jin}, \citenamefont {Lippert},\ and\ \citenamefont {{De
  Raedt}}}]{Michielsen2017BenchmarkingQC}%
  \BibitemOpen
  \bibfield  {author} {\bibinfo {author} {\bibfnamefont {K.}~\bibnamefont
  {Michielsen}}, \bibinfo {author} {\bibfnamefont {M.}~\bibnamefont {Nocon}},
  \bibinfo {author} {\bibfnamefont {D.}~\bibnamefont {Willsch}}, \bibinfo
  {author} {\bibfnamefont {F.}~\bibnamefont {Jin}}, \bibinfo {author}
  {\bibfnamefont {{\relax Th}.}~\bibnamefont {Lippert}}, \ and\ \bibinfo
  {author} {\bibfnamefont {H.}~\bibnamefont {{De Raedt}}},\ }\bibfield  {title}
  {\enquote {\bibinfo {title} {Benchmarking gate-based quantum computers},}\
  }\href {\doibase 10.1016/j.cpc.2017.06.011} {\bibfield  {journal} {\bibinfo
  {journal} {Comput. Phys. Commun.}\ }\textbf {\bibinfo {volume} {220}},\
  \bibinfo {pages} {44 } (\bibinfo {year} {2017})}\BibitemShut {NoStop}%
\bibitem [{\citenamefont {Shor}(1996)}]{Shor1996FaultTolerantQC}%
  \BibitemOpen
  \bibfield  {author} {\bibinfo {author} {\bibfnamefont {P.~W.}\ \bibnamefont
  {Shor}},\ }\bibfield  {title} {\enquote {\bibinfo {title} {Fault-tolerant
  quantum computation},}\ }in\ \href {\doibase 10.1109/SFCS.1996.548464} {\emph
  {\bibinfo {booktitle} {Proceedings of 37th Conference on Foundations of
  Computer Science}}}\ (\bibinfo {year} {1996})\ pp.\ \bibinfo {pages}
  {56--65}\BibitemShut {NoStop}%
\bibitem [{\citenamefont {Gottesman}(1998)}]{Gottesman1998TheoryFTQC}%
  \BibitemOpen
  \bibfield  {author} {\bibinfo {author} {\bibfnamefont {D.}~\bibnamefont
  {Gottesman}},\ }\bibfield  {title} {\enquote {\bibinfo {title} {Theory of
  fault-tolerant quantum computation},}\ }\href {\doibase
  10.1103/PhysRevA.57.127} {\bibfield  {journal} {\bibinfo  {journal} {Phys.
  Rev. A}\ }\textbf {\bibinfo {volume} {57}},\ \bibinfo {pages} {127} (\bibinfo
  {year} {1998})}\BibitemShut {NoStop}%
\bibitem [{\citenamefont {Campbell}\ \emph {et~al.}(2017)\citenamefont
  {Campbell}, \citenamefont {Terhal},\ and\ \citenamefont
  {Vuillot}}]{Campbell2017RoadsTowardsFTQC}%
  \BibitemOpen
  \bibfield  {author} {\bibinfo {author} {\bibfnamefont {E.~T.}\ \bibnamefont
  {Campbell}}, \bibinfo {author} {\bibfnamefont {B.~M.}\ \bibnamefont
  {Terhal}}, \ and\ \bibinfo {author} {\bibfnamefont {C.}~\bibnamefont
  {Vuillot}},\ }\bibfield  {title} {\enquote {\bibinfo {title} {Roads towards
  fault-tolerant universal quantum computation},}\ }\href
  {http://dx.doi.org/10.1038/nature23460} {\bibfield  {journal} {\bibinfo
  {journal} {Nature}\ }\textbf {\bibinfo {volume} {549}},\ \bibinfo {pages}
  {172} (\bibinfo {year} {2017})}\BibitemShut {NoStop}%
\bibitem [{\citenamefont {Takita}\ \emph {et~al.}(2016)\citenamefont {Takita},
  \citenamefont {C\'orcoles}, \citenamefont {Magesan}, \citenamefont {Abdo},
  \citenamefont {Brink}, \citenamefont {Cross}, \citenamefont {Chow},\ and\
  \citenamefont {Gambetta}}]{takita2016demonstration}%
  \BibitemOpen
  \bibfield  {author} {\bibinfo {author} {\bibfnamefont {M.}~\bibnamefont
  {Takita}}, \bibinfo {author} {\bibfnamefont {A.~D.}\ \bibnamefont
  {C\'orcoles}}, \bibinfo {author} {\bibfnamefont {E.}~\bibnamefont {Magesan}},
  \bibinfo {author} {\bibfnamefont {B.}~\bibnamefont {Abdo}}, \bibinfo {author}
  {\bibfnamefont {M.}~\bibnamefont {Brink}}, \bibinfo {author} {\bibfnamefont
  {A.}~\bibnamefont {Cross}}, \bibinfo {author} {\bibfnamefont {J.~M.}\
  \bibnamefont {Chow}}, \ and\ \bibinfo {author} {\bibfnamefont {J.~M.}\
  \bibnamefont {Gambetta}},\ }\bibfield  {title} {\enquote {\bibinfo {title}
  {Demonstration of weight-four parity measurements in the surface code
  architecture},}\ }\href {\doibase 10.1103/PhysRevLett.117.210505} {\bibfield
  {journal} {\bibinfo  {journal} {Phys. Rev. Lett.}\ }\textbf {\bibinfo
  {volume} {117}},\ \bibinfo {pages} {210505} (\bibinfo {year}
  {2016})}\BibitemShut {NoStop}%
\bibitem [{\citenamefont {Kelly}\ \emph {et~al.}(2015)\citenamefont {Kelly},
  \citenamefont {Barends}, \citenamefont {Fowler}, \citenamefont {Megrant},
  \citenamefont {Jeffrey}, \citenamefont {White}, \citenamefont {Sank},
  \citenamefont {Mutus}, \citenamefont {Campbell}, \citenamefont {Chen},
  \citenamefont {Chen}, \citenamefont {Chiaro}, \citenamefont {Dunsworth},
  \citenamefont {Hoi}, \citenamefont {Neill}, \citenamefont {O'Malley},
  \citenamefont {Quintana}, \citenamefont {Roushan}, \citenamefont
  {Vainsencher}, \citenamefont {Wenner}, \citenamefont {Cleland},\ and\
  \citenamefont {Martinis}}]{kelly2015statepreservation9qubits}%
  \BibitemOpen
  \bibfield  {author} {\bibinfo {author} {\bibfnamefont {J.}~\bibnamefont
  {Kelly}}, \bibinfo {author} {\bibfnamefont {R.}~\bibnamefont {Barends}},
  \bibinfo {author} {\bibfnamefont {A.~G.}\ \bibnamefont {Fowler}}, \bibinfo
  {author} {\bibfnamefont {A.}~\bibnamefont {Megrant}}, \bibinfo {author}
  {\bibfnamefont {E.}~\bibnamefont {Jeffrey}}, \bibinfo {author} {\bibfnamefont
  {T.~C.}\ \bibnamefont {White}}, \bibinfo {author} {\bibfnamefont
  {D.}~\bibnamefont {Sank}}, \bibinfo {author} {\bibfnamefont {J.~Y.}\
  \bibnamefont {Mutus}}, \bibinfo {author} {\bibfnamefont {B.}~\bibnamefont
  {Campbell}}, \bibinfo {author} {\bibfnamefont {Y.}~\bibnamefont {Chen}},
  \bibinfo {author} {\bibfnamefont {Z.}~\bibnamefont {Chen}}, \bibinfo {author}
  {\bibfnamefont {B.}~\bibnamefont {Chiaro}}, \bibinfo {author} {\bibfnamefont
  {A.}~\bibnamefont {Dunsworth}}, \bibinfo {author} {\bibfnamefont {I.-C.}\
  \bibnamefont {Hoi}}, \bibinfo {author} {\bibfnamefont {C.}~\bibnamefont
  {Neill}}, \bibinfo {author} {\bibfnamefont {P.~J.~J.}\ \bibnamefont
  {O'Malley}}, \bibinfo {author} {\bibfnamefont {C.}~\bibnamefont {Quintana}},
  \bibinfo {author} {\bibfnamefont {P.}~\bibnamefont {Roushan}}, \bibinfo
  {author} {\bibfnamefont {A.}~\bibnamefont {Vainsencher}}, \bibinfo {author}
  {\bibfnamefont {J.}~\bibnamefont {Wenner}}, \bibinfo {author} {\bibfnamefont
  {A.~N.}\ \bibnamefont {Cleland}}, \ and\ \bibinfo {author} {\bibfnamefont
  {J.~M.}\ \bibnamefont {Martinis}},\ }\bibfield  {title} {\enquote {\bibinfo
  {title} {State preservation by repetitive error detection in a
  superconducting quantum circuit},}\ }\href
  {http://dx.doi.org/10.1038/nature14270} {\bibfield  {journal} {\bibinfo
  {journal} {Nature}\ }\textbf {\bibinfo {volume} {519}},\ \bibinfo {pages}
  {66} (\bibinfo {year} {2015})}\BibitemShut {NoStop}%
\bibitem [{\citenamefont {Chow}\ \emph {et~al.}(2014)\citenamefont {Chow},
  \citenamefont {Gambetta}, \citenamefont {Magesan}, \citenamefont {Abraham},
  \citenamefont {Cross}, \citenamefont {Johnson}, \citenamefont {Masluk},
  \citenamefont {Ryan}, \citenamefont {Smolin}, \citenamefont {Srinivasan},\
  and\ \citenamefont {Steffen}}]{chow2014implementingastrand}%
  \BibitemOpen
  \bibfield  {author} {\bibinfo {author} {\bibfnamefont {J.~M.}\ \bibnamefont
  {Chow}}, \bibinfo {author} {\bibfnamefont {J.~M.}\ \bibnamefont {Gambetta}},
  \bibinfo {author} {\bibfnamefont {E.}~\bibnamefont {Magesan}}, \bibinfo
  {author} {\bibfnamefont {D.~W.}\ \bibnamefont {Abraham}}, \bibinfo {author}
  {\bibfnamefont {A.~W.}\ \bibnamefont {Cross}}, \bibinfo {author}
  {\bibfnamefont {B.~R.}\ \bibnamefont {Johnson}}, \bibinfo {author}
  {\bibfnamefont {N.~A.}\ \bibnamefont {Masluk}}, \bibinfo {author}
  {\bibfnamefont {C.~A.}\ \bibnamefont {Ryan}}, \bibinfo {author}
  {\bibfnamefont {J.~A.}\ \bibnamefont {Smolin}}, \bibinfo {author}
  {\bibfnamefont {S.~J.}\ \bibnamefont {Srinivasan}}, \ and\ \bibinfo {author}
  {\bibfnamefont {M.}~\bibnamefont {Steffen}},\ }\bibfield  {title} {\enquote
  {\bibinfo {title} {Implementing a strand of a scalable fault-tolerant quantum
  computing fabric},}\ }\href {\doibase 10.1038/ncomms5015} {\bibfield
  {journal} {\bibinfo  {journal} {Nat. Commun.}\ }\textbf {\bibinfo {volume}
  {5}},\ \bibinfo {pages} {4015} (\bibinfo {year} {2014})}\BibitemShut
  {NoStop}%
\bibitem [{\citenamefont {C\'orcoles}\ \emph {et~al.}(2015)\citenamefont
  {C\'orcoles}, \citenamefont {Magesan}, \citenamefont {Srinivasan},
  \citenamefont {Cross}, \citenamefont {Steffen}, \citenamefont {Gambetta},\
  and\ \citenamefont {Chow}}]{corcoles2015demonstration}%
  \BibitemOpen
  \bibfield  {author} {\bibinfo {author} {\bibfnamefont {A.}~\bibnamefont
  {C\'orcoles}}, \bibinfo {author} {\bibfnamefont {E.}~\bibnamefont {Magesan}},
  \bibinfo {author} {\bibfnamefont {S.~J.}\ \bibnamefont {Srinivasan}},
  \bibinfo {author} {\bibfnamefont {A.~W.}\ \bibnamefont {Cross}}, \bibinfo
  {author} {\bibfnamefont {M.}~\bibnamefont {Steffen}}, \bibinfo {author}
  {\bibfnamefont {J.~M.}\ \bibnamefont {Gambetta}}, \ and\ \bibinfo {author}
  {\bibfnamefont {J.~M.}\ \bibnamefont {Chow}},\ }\bibfield  {title} {\enquote
  {\bibinfo {title} {Demonstration of a quantum error detection code using a
  square lattice of four superconducting qubits},}\ }\href
  {http://dx.doi.org/10.1038/ncomms7979} {\bibfield  {journal} {\bibinfo
  {journal} {Nat. Commun.}\ }\textbf {\bibinfo {volume} {6}},\ \bibinfo {pages}
  {6979} (\bibinfo {year} {2015})}\BibitemShut {NoStop}%
\bibitem [{\citenamefont {Rist{\`e}}\ \emph {et~al.}(2015)\citenamefont
  {Rist{\`e}}, \citenamefont {Poletto}, \citenamefont {Huang}, \citenamefont
  {Bruno}, \citenamefont {Vesterinen}, \citenamefont {Saira},\ and\
  \citenamefont {DiCarlo}}]{riste2015detecting}%
  \BibitemOpen
  \bibfield  {author} {\bibinfo {author} {\bibfnamefont {D.}~\bibnamefont
  {Rist{\`e}}}, \bibinfo {author} {\bibfnamefont {S.}~\bibnamefont {Poletto}},
  \bibinfo {author} {\bibfnamefont {M.-Z.}\ \bibnamefont {Huang}}, \bibinfo
  {author} {\bibfnamefont {A.}~\bibnamefont {Bruno}}, \bibinfo {author}
  {\bibfnamefont {V.}~\bibnamefont {Vesterinen}}, \bibinfo {author}
  {\bibfnamefont {O.-P.}\ \bibnamefont {Saira}}, \ and\ \bibinfo {author}
  {\bibfnamefont {L.}~\bibnamefont {DiCarlo}},\ }\bibfield  {title} {\enquote
  {\bibinfo {title} {Detecting bit-flip errors in a logical qubit using
  stabilizer measurements},}\ }\href {\doibase 10.1038/ncomms7983} {\bibfield
  {journal} {\bibinfo  {journal} {Nat. Commun.}\ }\textbf {\bibinfo {volume}
  {6}},\ \bibinfo {pages} {6983} (\bibinfo {year} {2015})}\BibitemShut
  {NoStop}%
\bibitem [{\citenamefont
  {Gottesman}(2016)}]{Gottesman2016quantumfaulttolerance}%
  \BibitemOpen
  \bibfield  {author} {\bibinfo {author} {\bibfnamefont {D.}~\bibnamefont
  {Gottesman}},\ }\href@noop {} {\enquote {\bibinfo {title} {Quantum fault
  tolerance in small experiments},}\ } (\bibinfo {year} {2016}),\ \Eprint
  {http://arxiv.org/abs/1610.03507} {arXiv:1610.03507} \BibitemShut {NoStop}%
\bibitem [{\citenamefont {Leung}\ \emph {et~al.}(1997)\citenamefont {Leung},
  \citenamefont {Nielsen}, \citenamefont {Chuang},\ and\ \citenamefont
  {Yamamoto}}]{Leung1997fourqubitcode}%
  \BibitemOpen
  \bibfield  {author} {\bibinfo {author} {\bibfnamefont {D.~W.}\ \bibnamefont
  {Leung}}, \bibinfo {author} {\bibfnamefont {M.~A.}\ \bibnamefont {Nielsen}},
  \bibinfo {author} {\bibfnamefont {I.~L.}\ \bibnamefont {Chuang}}, \ and\
  \bibinfo {author} {\bibfnamefont {Y.}~\bibnamefont {Yamamoto}},\ }\bibfield
  {title} {\enquote {\bibinfo {title} {Approximate quantum error correction can
  lead to better codes},}\ }\href {\doibase 10.1103/PhysRevA.56.2567}
  {\bibfield  {journal} {\bibinfo  {journal} {Phys. Rev. A}\ }\textbf {\bibinfo
  {volume} {56}},\ \bibinfo {pages} {2567} (\bibinfo {year}
  {1997})}\BibitemShut {NoStop}%
\bibitem [{\citenamefont {Vaidman}\ \emph {et~al.}(1996)\citenamefont
  {Vaidman}, \citenamefont {Goldenberg},\ and\ \citenamefont
  {Wiesner}}]{Vaidman1996fourqubitcode}%
  \BibitemOpen
  \bibfield  {author} {\bibinfo {author} {\bibfnamefont {L.}~\bibnamefont
  {Vaidman}}, \bibinfo {author} {\bibfnamefont {L.}~\bibnamefont {Goldenberg}},
  \ and\ \bibinfo {author} {\bibfnamefont {S.}~\bibnamefont {Wiesner}},\
  }\bibfield  {title} {\enquote {\bibinfo {title} {Error prevention scheme with
  four particles},}\ }\href {\doibase 10.1103/PhysRevA.54.R1745} {\bibfield
  {journal} {\bibinfo  {journal} {Phys. Rev. A}\ }\textbf {\bibinfo {volume}
  {54}},\ \bibinfo {pages} {R1745} (\bibinfo {year} {1996})}\BibitemShut
  {NoStop}%
\bibitem [{\citenamefont {Grassl}\ \emph {et~al.}(1997)\citenamefont {Grassl},
  \citenamefont {Beth},\ and\ \citenamefont
  {Pellizzari}}]{Grassl1997fourqubitcode}%
  \BibitemOpen
  \bibfield  {author} {\bibinfo {author} {\bibfnamefont {M.}~\bibnamefont
  {Grassl}}, \bibinfo {author} {\bibfnamefont {T.}~\bibnamefont {Beth}}, \ and\
  \bibinfo {author} {\bibfnamefont {T.}~\bibnamefont {Pellizzari}},\ }\bibfield
   {title} {\enquote {\bibinfo {title} {Codes for the quantum erasure
  channel},}\ }\href {\doibase 10.1103/PhysRevA.56.33} {\bibfield  {journal}
  {\bibinfo  {journal} {Phys. Rev. A}\ }\textbf {\bibinfo {volume} {56}},\
  \bibinfo {pages} {33} (\bibinfo {year} {1997})}\BibitemShut {NoStop}%
\bibitem [{\citenamefont {Linke}\ \emph {et~al.}(2017)\citenamefont {Linke},
  \citenamefont {Gutierrez}, \citenamefont {Landsman}, \citenamefont {Figgatt},
  \citenamefont {Debnath}, \citenamefont {Brown},\ and\ \citenamefont
  {Monroe}}]{Linke2016FTIonTrapQubits}%
  \BibitemOpen
  \bibfield  {author} {\bibinfo {author} {\bibfnamefont {N.~M.}\ \bibnamefont
  {Linke}}, \bibinfo {author} {\bibfnamefont {M.}~\bibnamefont {Gutierrez}},
  \bibinfo {author} {\bibfnamefont {K.~A.}\ \bibnamefont {Landsman}}, \bibinfo
  {author} {\bibfnamefont {C.}~\bibnamefont {Figgatt}}, \bibinfo {author}
  {\bibfnamefont {S.}~\bibnamefont {Debnath}}, \bibinfo {author} {\bibfnamefont
  {K.~R.}\ \bibnamefont {Brown}}, \ and\ \bibinfo {author} {\bibfnamefont
  {C.}~\bibnamefont {Monroe}},\ }\bibfield  {title} {\enquote {\bibinfo {title}
  {Fault-tolerant quantum error detection},}\ }\href {\doibase
  10.1126/sciadv.1701074} {\bibfield  {journal} {\bibinfo  {journal} {Sci.
  Adv.}\ }\textbf {\bibinfo {volume} {3}},\ \bibinfo {pages} {e1701074}
  (\bibinfo {year} {2017})}\BibitemShut {NoStop}%
\bibitem [{\citenamefont {Vuillot}(2018)}]{Vuillot2017ErrorDetectionIBM}%
  \BibitemOpen
  \bibfield  {author} {\bibinfo {author} {\bibfnamefont {C.}~\bibnamefont
  {Vuillot}},\ }\bibfield  {title} {\enquote {\bibinfo {title} {Is error
  detection helpful on {IBM} 5{Q} chips?}}\ }\href@noop {} {\bibfield
  {journal} {\bibinfo  {journal} {Quantum Inf. Comput.}\ }\textbf {\bibinfo
  {volume} {18}},\ \bibinfo {pages} {0949} (\bibinfo {year} {2018})},\ \Eprint
  {http://arxiv.org/abs/1705.08957v2} {arXiv:1705.08957v2} \BibitemShut
  {NoStop}%
\bibitem [{\citenamefont {Takita}\ \emph {et~al.}(2017)\citenamefont {Takita},
  \citenamefont {Cross}, \citenamefont {C\'orcoles}, \citenamefont {Chow},\
  and\ \citenamefont {Gambetta}}]{Takita2017faultTolerantStatePreparation}%
  \BibitemOpen
  \bibfield  {author} {\bibinfo {author} {\bibfnamefont {M.}~\bibnamefont
  {Takita}}, \bibinfo {author} {\bibfnamefont {A.~W.}\ \bibnamefont {Cross}},
  \bibinfo {author} {\bibfnamefont {A.~D.}\ \bibnamefont {C\'orcoles}},
  \bibinfo {author} {\bibfnamefont {J.~M.}\ \bibnamefont {Chow}}, \ and\
  \bibinfo {author} {\bibfnamefont {J.~M.}\ \bibnamefont {Gambetta}},\
  }\bibfield  {title} {\enquote {\bibinfo {title} {Experimental demonstration
  of fault-tolerant state preparation with superconducting qubits},}\ }\href
  {\doibase 10.1103/PhysRevLett.119.180501} {\bibfield  {journal} {\bibinfo
  {journal} {Phys. Rev. Lett.}\ }\textbf {\bibinfo {volume} {119}},\ \bibinfo
  {pages} {180501} (\bibinfo {year} {2017})}\BibitemShut {NoStop}%
\bibitem [{\citenamefont {Harper}\ and\ \citenamefont
  {Flammia}(2018)}]{HarperFlammia2018FaultToleranceInTheIBMQ}%
  \BibitemOpen
  \bibfield  {author} {\bibinfo {author} {\bibfnamefont {R.}~\bibnamefont
  {Harper}}\ and\ \bibinfo {author} {\bibfnamefont {S.}~\bibnamefont
  {Flammia}},\ }\href@noop {} {\enquote {\bibinfo {title} {Fault tolerance in
  the {IBM Q Experience}},}\ } (\bibinfo {year} {2018}),\ \Eprint
  {http://arxiv.org/abs/1806.02359} {arXiv:1806.02359} \BibitemShut {NoStop}%
\bibitem [{\citenamefont {Jin}\ \emph {et~al.}(2010)\citenamefont {Jin},
  \citenamefont {{De Raedt}}, \citenamefont {Yuan}, \citenamefont {Katsnelson},
  \citenamefont {Miyashita},\ and\ \citenamefont
  {Michielsen}}]{Jin2010approachtoequilibrium}%
  \BibitemOpen
  \bibfield  {author} {\bibinfo {author} {\bibfnamefont {F.}~\bibnamefont
  {Jin}}, \bibinfo {author} {\bibfnamefont {H.}~\bibnamefont {{De Raedt}}},
  \bibinfo {author} {\bibfnamefont {S.}~\bibnamefont {Yuan}}, \bibinfo {author}
  {\bibfnamefont {M.~I.}\ \bibnamefont {Katsnelson}}, \bibinfo {author}
  {\bibfnamefont {S.}~\bibnamefont {Miyashita}}, \ and\ \bibinfo {author}
  {\bibfnamefont {K.}~\bibnamefont {Michielsen}},\ }\bibfield  {title}
  {\enquote {\bibinfo {title} {Approach to equilibrium in nano-scale systems at
  finite temperature},}\ }\href {\doibase 10.1143/JPSJ.79.124005} {\bibfield
  {journal} {\bibinfo  {journal} {J. Phys. Soc. Jpn.}\ }\textbf {\bibinfo
  {volume} {79}},\ \bibinfo {pages} {124005} (\bibinfo {year}
  {2010})}\BibitemShut {NoStop}%
\bibitem [{\citenamefont {Zhao}\ \emph {et~al.}(2016)\citenamefont {Zhao},
  \citenamefont {{De Raedt}}, \citenamefont {Miyashita}, \citenamefont {Jin},\
  and\ \citenamefont {Michielsen}}]{zhao2016masterequation}%
  \BibitemOpen
  \bibfield  {author} {\bibinfo {author} {\bibfnamefont {P.}~\bibnamefont
  {Zhao}}, \bibinfo {author} {\bibfnamefont {H.}~\bibnamefont {{De Raedt}}},
  \bibinfo {author} {\bibfnamefont {S.}~\bibnamefont {Miyashita}}, \bibinfo
  {author} {\bibfnamefont {F.}~\bibnamefont {Jin}}, \ and\ \bibinfo {author}
  {\bibfnamefont {K.}~\bibnamefont {Michielsen}},\ }\bibfield  {title}
  {\enquote {\bibinfo {title} {Dynamics of open quantum spin systems: An
  assessment of the quantum master equation approach},}\ }\href {\doibase
  10.1103/PhysRevE.94.022126} {\bibfield  {journal} {\bibinfo  {journal} {Phys.
  Rev. E}\ }\textbf {\bibinfo {volume} {94}},\ \bibinfo {pages} {022126}
  (\bibinfo {year} {2016})}\BibitemShut {NoStop}%
\bibitem [{\citenamefont {M\"uller}\ \emph {et~al.}(2015)\citenamefont
  {M\"uller}, \citenamefont {Lisenfeld}, \citenamefont {Shnirman},\ and\
  \citenamefont {Poletto}}]{Mueller2015twoleveldefects}%
  \BibitemOpen
  \bibfield  {author} {\bibinfo {author} {\bibfnamefont {C.}~\bibnamefont
  {M\"uller}}, \bibinfo {author} {\bibfnamefont {J.}~\bibnamefont {Lisenfeld}},
  \bibinfo {author} {\bibfnamefont {A.}~\bibnamefont {Shnirman}}, \ and\
  \bibinfo {author} {\bibfnamefont {S.}~\bibnamefont {Poletto}},\ }\bibfield
  {title} {\enquote {\bibinfo {title} {Interacting two-level defects as sources
  of fluctuating high-frequency noise in superconducting circuits},}\ }\href
  {\doibase 10.1103/PhysRevB.92.035442} {\bibfield  {journal} {\bibinfo
  {journal} {Phys. Rev. B}\ }\textbf {\bibinfo {volume} {92}},\ \bibinfo
  {pages} {035442} (\bibinfo {year} {2015})}\BibitemShut {NoStop}%
\bibitem [{\citenamefont {Magesan}\ \emph {et~al.}(2013)\citenamefont
  {Magesan}, \citenamefont {Puzzuoli}, \citenamefont {Granade},\ and\
  \citenamefont {Cory}}]{Magesan2013ModelingQuantumNoise}%
  \BibitemOpen
  \bibfield  {author} {\bibinfo {author} {\bibfnamefont {E.}~\bibnamefont
  {Magesan}}, \bibinfo {author} {\bibfnamefont {D.}~\bibnamefont {Puzzuoli}},
  \bibinfo {author} {\bibfnamefont {C.~E.}\ \bibnamefont {Granade}}, \ and\
  \bibinfo {author} {\bibfnamefont {D.~G.}\ \bibnamefont {Cory}},\ }\bibfield
  {title} {\enquote {\bibinfo {title} {Modeling quantum noise for efficient
  testing of fault-tolerant circuits},}\ }\href {\doibase
  10.1103/PhysRevA.87.012324} {\bibfield  {journal} {\bibinfo  {journal} {Phys.
  Rev. A}\ }\textbf {\bibinfo {volume} {87}},\ \bibinfo {pages} {012324}
  (\bibinfo {year} {2013})}\BibitemShut {NoStop}%
\bibitem [{\citenamefont {Puzzuoli}\ \emph {et~al.}(2014)\citenamefont
  {Puzzuoli}, \citenamefont {Granade}, \citenamefont {Haas}, \citenamefont
  {Criger}, \citenamefont {Magesan},\ and\ \citenamefont
  {Cory}}]{Puzzuoli2014tractablesimulation}%
  \BibitemOpen
  \bibfield  {author} {\bibinfo {author} {\bibfnamefont {D.}~\bibnamefont
  {Puzzuoli}}, \bibinfo {author} {\bibfnamefont {C.}~\bibnamefont {Granade}},
  \bibinfo {author} {\bibfnamefont {H.}~\bibnamefont {Haas}}, \bibinfo {author}
  {\bibfnamefont {B.}~\bibnamefont {Criger}}, \bibinfo {author} {\bibfnamefont
  {E.}~\bibnamefont {Magesan}}, \ and\ \bibinfo {author} {\bibfnamefont
  {D.~G.}\ \bibnamefont {Cory}},\ }\bibfield  {title} {\enquote {\bibinfo
  {title} {Tractable simulation of error correction with honest approximations
  to realistic fault models},}\ }\href {\doibase 10.1103/PhysRevA.89.022306}
  {\bibfield  {journal} {\bibinfo  {journal} {Phys. Rev. A}\ }\textbf {\bibinfo
  {volume} {89}},\ \bibinfo {pages} {022306} (\bibinfo {year}
  {2014})}\BibitemShut {NoStop}%
\bibitem [{\citenamefont {Iyer}\ and\ \citenamefont
  {Poulin}(2018)}]{Iyer2017smallQCneededforFT}%
  \BibitemOpen
  \bibfield  {author} {\bibinfo {author} {\bibfnamefont {P.~S.}\ \bibnamefont
  {Iyer}}\ and\ \bibinfo {author} {\bibfnamefont {D.}~\bibnamefont {Poulin}},\
  }\bibfield  {title} {\enquote {\bibinfo {title} {A small quantum computer is
  needed to optimize fault-tolerant protocols},}\ }\href
  {http://stacks.iop.org/2058-9565/3/i=3/a=030504} {\bibfield  {journal}
  {\bibinfo  {journal} {Quantum Sci. Technol.}\ }\textbf {\bibinfo {volume}
  {3}},\ \bibinfo {pages} {030504} (\bibinfo {year} {2018})}\BibitemShut
  {NoStop}%
\bibitem [{\citenamefont {Nielsen}\ and\ \citenamefont
  {Chuang}(2011)}]{NielsenChuang}%
  \BibitemOpen
  \bibfield  {author} {\bibinfo {author} {\bibfnamefont {M.~A.}\ \bibnamefont
  {Nielsen}}\ and\ \bibinfo {author} {\bibfnamefont {I.~L.}\ \bibnamefont
  {Chuang}},\ }\href@noop {} {\emph {\bibinfo {title} {Quantum Computation and
  Quantum Information: 10th Anniversary Edition}}}\ (\bibinfo  {publisher}
  {Cambridge University Press},\ \bibinfo {address} {New York},\ \bibinfo
  {year} {2011})\BibitemShut {NoStop}%
\bibitem [{\citenamefont {Terhal}\ and\ \citenamefont
  {Burkard}(2005)}]{Terhal2005ftqcForLocalNonmarkovianNoise}%
  \BibitemOpen
  \bibfield  {author} {\bibinfo {author} {\bibfnamefont {B.~M.}\ \bibnamefont
  {Terhal}}\ and\ \bibinfo {author} {\bibfnamefont {G.}~\bibnamefont
  {Burkard}},\ }\bibfield  {title} {\enquote {\bibinfo {title} {Fault-tolerant
  quantum computation for local non-markovian noise},}\ }\href {\doibase
  10.1103/PhysRevA.71.012336} {\bibfield  {journal} {\bibinfo  {journal} {Phys.
  Rev. A}\ }\textbf {\bibinfo {volume} {71}},\ \bibinfo {pages} {012336}
  (\bibinfo {year} {2005})}\BibitemShut {NoStop}%
\bibitem [{\citenamefont {Aliferis}\ \emph {et~al.}(2006)\citenamefont
  {Aliferis}, \citenamefont {Gottesman},\ and\ \citenamefont
  {Preskill}}]{aliferis2006extendedrectangles}%
  \BibitemOpen
  \bibfield  {author} {\bibinfo {author} {\bibfnamefont {P.}~\bibnamefont
  {Aliferis}}, \bibinfo {author} {\bibfnamefont {D.}~\bibnamefont {Gottesman}},
  \ and\ \bibinfo {author} {\bibfnamefont {J.}~\bibnamefont {Preskill}},\
  }\bibfield  {title} {\enquote {\bibinfo {title} {Quantum accuracy threshold
  for concatenated distance-3 codes},}\ }\href
  {http://dl.acm.org/citation.cfm?id=2011665.2011666} {\bibfield  {journal}
  {\bibinfo  {journal} {Quantum Inf. Comput.}\ }\textbf {\bibinfo {volume}
  {6}},\ \bibinfo {pages} {97} (\bibinfo {year} {2006})}\BibitemShut {NoStop}%
\bibitem [{\citenamefont {Aliferis}\ and\ \citenamefont
  {Terhal}(2007)}]{aliferis2007FTQCwithLeakage}%
  \BibitemOpen
  \bibfield  {author} {\bibinfo {author} {\bibfnamefont {P.}~\bibnamefont
  {Aliferis}}\ and\ \bibinfo {author} {\bibfnamefont {B.~M.}\ \bibnamefont
  {Terhal}},\ }\bibfield  {title} {\enquote {\bibinfo {title} {Fault-tolerant
  quantum computation for local leakage faults},}\ }\href
  {http://dl.acm.org/citation.cfm?id=2011706.2011715} {\bibfield  {journal}
  {\bibinfo  {journal} {Quantum Inf. Comput.}\ }\textbf {\bibinfo {volume}
  {7}},\ \bibinfo {pages} {139} (\bibinfo {year} {2007})}\BibitemShut {NoStop}%
\bibitem [{\citenamefont {Aharonov}\ and\ \citenamefont
  {Ben-Or}(2008)}]{aharonov2008thresholdtheorem}%
  \BibitemOpen
  \bibfield  {author} {\bibinfo {author} {\bibfnamefont {D.}~\bibnamefont
  {Aharonov}}\ and\ \bibinfo {author} {\bibfnamefont {M.}~\bibnamefont
  {Ben-Or}},\ }\bibfield  {title} {\enquote {\bibinfo {title} {Fault-tolerant
  quantum computation with constant error rate},}\ }\href {\doibase
  10.1137/S0097539799359385} {\bibfield  {journal} {\bibinfo  {journal} {SIAM
  J. Comput.}\ }\textbf {\bibinfo {volume} {38}},\ \bibinfo {pages} {1207}
  (\bibinfo {year} {2008})}\BibitemShut {NoStop}%
\bibitem [{\citenamefont {Ng}\ and\ \citenamefont
  {Preskill}(2009)}]{ng2009FTQCversusGaussianNoise}%
  \BibitemOpen
  \bibfield  {author} {\bibinfo {author} {\bibfnamefont {H.~K.}\ \bibnamefont
  {Ng}}\ and\ \bibinfo {author} {\bibfnamefont {J.}~\bibnamefont {Preskill}},\
  }\bibfield  {title} {\enquote {\bibinfo {title} {Fault-tolerant quantum
  computation versus gaussian noise},}\ }\href {\doibase
  10.1103/PhysRevA.79.032318} {\bibfield  {journal} {\bibinfo  {journal} {Phys.
  Rev. A}\ }\textbf {\bibinfo {volume} {79}},\ \bibinfo {pages} {032318}
  (\bibinfo {year} {2009})}\BibitemShut {NoStop}%
\bibitem [{\citenamefont {Kitaev}(1997)}]{kitaev1997diamondnorm}%
  \BibitemOpen
  \bibfield  {author} {\bibinfo {author} {\bibfnamefont {A.~Y.}\ \bibnamefont
  {Kitaev}},\ }\bibfield  {title} {\enquote {\bibinfo {title} {Quantum
  computations: algorithms and error correction},}\ }\href
  {http://stacks.iop.org/0036-0279/52/i=6/a=R02} {\bibfield  {journal}
  {\bibinfo  {journal} {Russ. Math. Surveys}\ }\textbf {\bibinfo {volume}
  {52}},\ \bibinfo {pages} {1191} (\bibinfo {year} {1997})}\BibitemShut
  {NoStop}%
\bibitem [{\citenamefont {Sanders}\ \emph {et~al.}(2016)\citenamefont
  {Sanders}, \citenamefont {Wallman},\ and\ \citenamefont
  {Sanders}}]{Sanders2016ThresholdTheorem}%
  \BibitemOpen
  \bibfield  {author} {\bibinfo {author} {\bibfnamefont {Y.~R.}\ \bibnamefont
  {Sanders}}, \bibinfo {author} {\bibfnamefont {J.~J.}\ \bibnamefont
  {Wallman}}, \ and\ \bibinfo {author} {\bibfnamefont {B.~C.}\ \bibnamefont
  {Sanders}},\ }\bibfield  {title} {\enquote {\bibinfo {title} {Bounding
  quantum gate error rate based on reported average fidelity},}\ }\href
  {http://stacks.iop.org/1367-2630/18/i=1/a=012002} {\bibfield  {journal}
  {\bibinfo  {journal} {New J. Phys.}\ }\textbf {\bibinfo {volume} {18}},\
  \bibinfo {pages} {012002} (\bibinfo {year} {2016})}\BibitemShut {NoStop}%
\bibitem [{\citenamefont {Kueng}\ \emph {et~al.}(2016)\citenamefont {Kueng},
  \citenamefont {Long}, \citenamefont {Doherty},\ and\ \citenamefont
  {Flammia}}]{Kueng2016ComparingExperimentsToThreshold}%
  \BibitemOpen
  \bibfield  {author} {\bibinfo {author} {\bibfnamefont {R.}~\bibnamefont
  {Kueng}}, \bibinfo {author} {\bibfnamefont {D.~M.}\ \bibnamefont {Long}},
  \bibinfo {author} {\bibfnamefont {A.~C.}\ \bibnamefont {Doherty}}, \ and\
  \bibinfo {author} {\bibfnamefont {S.~T.}\ \bibnamefont {Flammia}},\
  }\bibfield  {title} {\enquote {\bibinfo {title} {Comparing experiments to the
  fault-tolerance threshold},}\ }\href {\doibase
  10.1103/PhysRevLett.117.170502} {\bibfield  {journal} {\bibinfo  {journal}
  {Phys. Rev. Lett.}\ }\textbf {\bibinfo {volume} {117}},\ \bibinfo {pages}
  {170502} (\bibinfo {year} {2016})}\BibitemShut {NoStop}%
\bibitem [{\citenamefont {Proctor}\ \emph {et~al.}(2017)\citenamefont
  {Proctor}, \citenamefont {Rudinger}, \citenamefont {Young}, \citenamefont
  {Sarovar},\ and\ \citenamefont
  {Blume-Kohout}}]{proctor2017RandomizedBenchmarking}%
  \BibitemOpen
  \bibfield  {author} {\bibinfo {author} {\bibfnamefont {T.}~\bibnamefont
  {Proctor}}, \bibinfo {author} {\bibfnamefont {K.}~\bibnamefont {Rudinger}},
  \bibinfo {author} {\bibfnamefont {K.}~\bibnamefont {Young}}, \bibinfo
  {author} {\bibfnamefont {M.}~\bibnamefont {Sarovar}}, \ and\ \bibinfo
  {author} {\bibfnamefont {R.}~\bibnamefont {Blume-Kohout}},\ }\bibfield
  {title} {\enquote {\bibinfo {title} {What randomized benchmarking actually
  measures},}\ }\href {\doibase 10.1103/PhysRevLett.119.130502} {\bibfield
  {journal} {\bibinfo  {journal} {Phys. Rev. Lett.}\ }\textbf {\bibinfo
  {volume} {119}},\ \bibinfo {pages} {130502} (\bibinfo {year}
  {2017})}\BibitemShut {NoStop}%
\bibitem [{\citenamefont {Barends}\ \emph {et~al.}(2013)\citenamefont
  {Barends}, \citenamefont {Kelly}, \citenamefont {Megrant}, \citenamefont
  {Sank}, \citenamefont {Jeffrey}, \citenamefont {Chen}, \citenamefont {Yin},
  \citenamefont {Chiaro}, \citenamefont {Mutus}, \citenamefont {Neill},
  \citenamefont {O'Malley}, \citenamefont {Roushan}, \citenamefont {Wenner},
  \citenamefont {White}, \citenamefont {Cleland},\ and\ \citenamefont
  {Martinis}}]{barendsMartinis2013xmoncoherence}%
  \BibitemOpen
  \bibfield  {author} {\bibinfo {author} {\bibfnamefont {R.}~\bibnamefont
  {Barends}}, \bibinfo {author} {\bibfnamefont {J.}~\bibnamefont {Kelly}},
  \bibinfo {author} {\bibfnamefont {A.}~\bibnamefont {Megrant}}, \bibinfo
  {author} {\bibfnamefont {D.}~\bibnamefont {Sank}}, \bibinfo {author}
  {\bibfnamefont {E.}~\bibnamefont {Jeffrey}}, \bibinfo {author} {\bibfnamefont
  {Y.}~\bibnamefont {Chen}}, \bibinfo {author} {\bibfnamefont {Y.}~\bibnamefont
  {Yin}}, \bibinfo {author} {\bibfnamefont {B.}~\bibnamefont {Chiaro}},
  \bibinfo {author} {\bibfnamefont {J.}~\bibnamefont {Mutus}}, \bibinfo
  {author} {\bibfnamefont {C.}~\bibnamefont {Neill}}, \bibinfo {author}
  {\bibfnamefont {P.}~\bibnamefont {O'Malley}}, \bibinfo {author}
  {\bibfnamefont {P.}~\bibnamefont {Roushan}}, \bibinfo {author} {\bibfnamefont
  {J.}~\bibnamefont {Wenner}}, \bibinfo {author} {\bibfnamefont {T.~C.}\
  \bibnamefont {White}}, \bibinfo {author} {\bibfnamefont {A.~N.}\ \bibnamefont
  {Cleland}}, \ and\ \bibinfo {author} {\bibfnamefont {J.~M.}\ \bibnamefont
  {Martinis}},\ }\bibfield  {title} {\enquote {\bibinfo {title} {Coherent
  josephson qubit suitable for scalable quantum integrated circuits},}\ }\href
  {\doibase 10.1103/PhysRevLett.111.080502} {\bibfield  {journal} {\bibinfo
  {journal} {Phys. Rev. Lett.}\ }\textbf {\bibinfo {volume} {111}},\ \bibinfo
  {pages} {080502} (\bibinfo {year} {2013})}\BibitemShut {NoStop}%
\bibitem [{\citenamefont {Wang}\ \emph {et~al.}(2015)\citenamefont {Wang},
  \citenamefont {Axline}, \citenamefont {Gao}, \citenamefont {Brecht},
  \citenamefont {Chu}, \citenamefont {Frunzio}, \citenamefont {Devoret},\ and\
  \citenamefont {Schoelkopf}}]{Wang2015dielectricloss}%
  \BibitemOpen
  \bibfield  {author} {\bibinfo {author} {\bibfnamefont {C.}~\bibnamefont
  {Wang}}, \bibinfo {author} {\bibfnamefont {C.}~\bibnamefont {Axline}},
  \bibinfo {author} {\bibfnamefont {Y.~Y.}\ \bibnamefont {Gao}}, \bibinfo
  {author} {\bibfnamefont {T.}~\bibnamefont {Brecht}}, \bibinfo {author}
  {\bibfnamefont {Y.}~\bibnamefont {Chu}}, \bibinfo {author} {\bibfnamefont
  {L.}~\bibnamefont {Frunzio}}, \bibinfo {author} {\bibfnamefont {M.~H.}\
  \bibnamefont {Devoret}}, \ and\ \bibinfo {author} {\bibfnamefont {R.~J.}\
  \bibnamefont {Schoelkopf}},\ }\bibfield  {title} {\enquote {\bibinfo {title}
  {Surface participation and dielectric loss in superconducting qubits},}\
  }\href {\doibase 10.1063/1.4934486} {\bibfield  {journal} {\bibinfo
  {journal} {Appl. Phys. Lett.}\ }\textbf {\bibinfo {volume} {107}},\ \bibinfo
  {pages} {162601} (\bibinfo {year} {2015})}\BibitemShut {NoStop}%
\bibitem [{\citenamefont {{De Raedt}}\ \emph {et~al.}(2018)\citenamefont {{De
  Raedt}}, \citenamefont {Jin}, \citenamefont {Willsch}, \citenamefont {Nocon},
  \citenamefont {Yoshioka}, \citenamefont {Ito}, \citenamefont {Yuan},\ and\
  \citenamefont {Michielsen}}]{DeRaedt2017MassivelyParallel}%
  \BibitemOpen
  \bibfield  {author} {\bibinfo {author} {\bibfnamefont {H.}~\bibnamefont {{De
  Raedt}}}, \bibinfo {author} {\bibfnamefont {F.}~\bibnamefont {Jin}}, \bibinfo
  {author} {\bibfnamefont {D.}~\bibnamefont {Willsch}}, \bibinfo {author}
  {\bibfnamefont {M.}~\bibnamefont {Nocon}}, \bibinfo {author} {\bibfnamefont
  {N.}~\bibnamefont {Yoshioka}}, \bibinfo {author} {\bibfnamefont
  {N.}~\bibnamefont {Ito}}, \bibinfo {author} {\bibfnamefont {S.}~\bibnamefont
  {Yuan}}, \ and\ \bibinfo {author} {\bibfnamefont {K.}~\bibnamefont
  {Michielsen}},\ }\href@noop {} {\enquote {\bibinfo {title} {Massively
  parallel quantum computer simulator, eleven years later},}\ } (\bibinfo
  {year} {2018}),\ \Eprint {http://arxiv.org/abs/1805.04708} {arXiv:1805.04708}
  \BibitemShut {NoStop}%
\bibitem [{\citenamefont {Tal-Ezer}\ and\ \citenamefont
  {Kosloff}(1984)}]{talezerkosloff1984chebyshev}%
  \BibitemOpen
  \bibfield  {author} {\bibinfo {author} {\bibfnamefont {H.}~\bibnamefont
  {Tal-Ezer}}\ and\ \bibinfo {author} {\bibfnamefont {R.}~\bibnamefont
  {Kosloff}},\ }\bibfield  {title} {\enquote {\bibinfo {title} {An accurate and
  efficient scheme for propagating the time dependent {S}chr{\"o}dinger
  equation},}\ }\href {\doibase 10.1063/1.448136} {\bibfield  {journal}
  {\bibinfo  {journal} {J. Chem. Phys.}\ }\textbf {\bibinfo {volume} {81}},\
  \bibinfo {pages} {3967} (\bibinfo {year} {1984})}\BibitemShut {NoStop}%
\bibitem [{\citenamefont {Dobrovitski}\ and\ \citenamefont
  {De~Raedt}(2003)}]{dobrovitski2003chebyshev}%
  \BibitemOpen
  \bibfield  {author} {\bibinfo {author} {\bibfnamefont {V.~V.}\ \bibnamefont
  {Dobrovitski}}\ and\ \bibinfo {author} {\bibfnamefont {H.~A.}\ \bibnamefont
  {De~Raedt}},\ }\bibfield  {title} {\enquote {\bibinfo {title} {Efficient
  scheme for numerical simulations of the spin-bath decoherence},}\ }\href
  {\doibase 10.1103/PhysRevE.67.056702} {\bibfield  {journal} {\bibinfo
  {journal} {Phys. Rev. E}\ }\textbf {\bibinfo {volume} {67}},\ \bibinfo
  {pages} {056702} (\bibinfo {year} {2003})}\BibitemShut {NoStop}%
\bibitem [{\citenamefont {{De Raedt}}\ \emph {et~al.}(2017)\citenamefont {{De
  Raedt}}, \citenamefont {Jin}, \citenamefont {Katsnelson},\ and\ \citenamefont
  {Michielsen}}]{DeRaedt2017relaxation}%
  \BibitemOpen
  \bibfield  {author} {\bibinfo {author} {\bibfnamefont {H.}~\bibnamefont {{De
  Raedt}}}, \bibinfo {author} {\bibfnamefont {F.}~\bibnamefont {Jin}}, \bibinfo
  {author} {\bibfnamefont {M.~I.}\ \bibnamefont {Katsnelson}}, \ and\ \bibinfo
  {author} {\bibfnamefont {K.}~\bibnamefont {Michielsen}},\ }\bibfield  {title}
  {\enquote {\bibinfo {title} {Relaxation, thermalization, and markovian
  dynamics of two spins coupled to a spin bath},}\ }\href {\doibase
  10.1103/PhysRevE.96.053306} {\bibfield  {journal} {\bibinfo  {journal} {Phys.
  Rev. E}\ }\textbf {\bibinfo {volume} {96}},\ \bibinfo {pages} {053306}
  (\bibinfo {year} {2017})}\BibitemShut {NoStop}%
\bibitem [{\citenamefont {Hams}\ and\ \citenamefont {{De
  Raedt}}(2000)}]{HamsDeRaedt2000RandomStateTechnology}%
  \BibitemOpen
  \bibfield  {author} {\bibinfo {author} {\bibfnamefont {A.}~\bibnamefont
  {Hams}}\ and\ \bibinfo {author} {\bibfnamefont {H.}~\bibnamefont {{De
  Raedt}}},\ }\bibfield  {title} {\enquote {\bibinfo {title} {Fast algorithm
  for finding the eigenvalue distribution of very large matrices},}\ }\href
  {\doibase 10.1103/PhysRevE.62.4365} {\bibfield  {journal} {\bibinfo
  {journal} {Phys. Rev. E}\ }\textbf {\bibinfo {volume} {62}},\ \bibinfo
  {pages} {4365} (\bibinfo {year} {2000})}\BibitemShut {NoStop}%
\bibitem [{\citenamefont {{De Raedt}}\ \emph {et~al.}(2012)\citenamefont {{De
  Raedt}}, \citenamefont {Barbara}, \citenamefont {Miyashita}, \citenamefont
  {Michielsen}, \citenamefont {Bertaina},\ and\ \citenamefont
  {Gambarelli}}]{deraedt2012rabioscillations}%
  \BibitemOpen
  \bibfield  {author} {\bibinfo {author} {\bibfnamefont {H.}~\bibnamefont {{De
  Raedt}}}, \bibinfo {author} {\bibfnamefont {B.}~\bibnamefont {Barbara}},
  \bibinfo {author} {\bibfnamefont {S.}~\bibnamefont {Miyashita}}, \bibinfo
  {author} {\bibfnamefont {K.}~\bibnamefont {Michielsen}}, \bibinfo {author}
  {\bibfnamefont {S.}~\bibnamefont {Bertaina}}, \ and\ \bibinfo {author}
  {\bibfnamefont {S.}~\bibnamefont {Gambarelli}},\ }\bibfield  {title}
  {\enquote {\bibinfo {title} {Quantum simulations and experiments on {R}abi
  oscillations of spin qubits: Intrinsic vs extrinsic damping},}\ }\href
  {\doibase 10.1103/PhysRevB.85.014408} {\bibfield  {journal} {\bibinfo
  {journal} {Phys. Rev. B}\ }\textbf {\bibinfo {volume} {85}},\ \bibinfo
  {pages} {014408} (\bibinfo {year} {2012})}\BibitemShut {NoStop}%
\bibitem [{\citenamefont {Sheldon}\ \emph
  {et~al.}(2016{\natexlab{b}})\citenamefont {Sheldon}, \citenamefont {Magesan},
  \citenamefont {Chow},\ and\ \citenamefont {Gambetta}}]{sheldon2016procedure}%
  \BibitemOpen
  \bibfield  {author} {\bibinfo {author} {\bibfnamefont {S.}~\bibnamefont
  {Sheldon}}, \bibinfo {author} {\bibfnamefont {E.}~\bibnamefont {Magesan}},
  \bibinfo {author} {\bibfnamefont {J.~M.}\ \bibnamefont {Chow}}, \ and\
  \bibinfo {author} {\bibfnamefont {J.~M.}\ \bibnamefont {Gambetta}},\
  }\bibfield  {title} {\enquote {\bibinfo {title} {Procedure for systematically
  tuning up cross-talk in the cross-resonance gate},}\ }\href {\doibase
  10.1103/PhysRevA.93.060302} {\bibfield  {journal} {\bibinfo  {journal} {Phys.
  Rev. A}\ }\textbf {\bibinfo {volume} {93}},\ \bibinfo {pages} {060302}
  (\bibinfo {year} {2016}{\natexlab{b}})}\BibitemShut {NoStop}%
\bibitem [{\citenamefont {Koch}\ \emph {et~al.}(2007)\citenamefont {Koch},
  \citenamefont {Yu}, \citenamefont {Gambetta}, \citenamefont {Houck},
  \citenamefont {Schuster}, \citenamefont {Majer}, \citenamefont {Blais},
  \citenamefont {Devoret}, \citenamefont {Girvin},\ and\ \citenamefont
  {Schoelkopf}}]{koch2007transmon}%
  \BibitemOpen
  \bibfield  {author} {\bibinfo {author} {\bibfnamefont {J.}~\bibnamefont
  {Koch}}, \bibinfo {author} {\bibfnamefont {T.~M.}\ \bibnamefont {Yu}},
  \bibinfo {author} {\bibfnamefont {J.}~\bibnamefont {Gambetta}}, \bibinfo
  {author} {\bibfnamefont {A.~A.}\ \bibnamefont {Houck}}, \bibinfo {author}
  {\bibfnamefont {D.~I.}\ \bibnamefont {Schuster}}, \bibinfo {author}
  {\bibfnamefont {J.}~\bibnamefont {Majer}}, \bibinfo {author} {\bibfnamefont
  {A.}~\bibnamefont {Blais}}, \bibinfo {author} {\bibfnamefont {M.~H.}\
  \bibnamefont {Devoret}}, \bibinfo {author} {\bibfnamefont {S.~M.}\
  \bibnamefont {Girvin}}, \ and\ \bibinfo {author} {\bibfnamefont {R.~J.}\
  \bibnamefont {Schoelkopf}},\ }\bibfield  {title} {\enquote {\bibinfo {title}
  {Charge-insensitive qubit design derived from the {C}ooper pair box},}\
  }\href {\doibase 10.1103/PhysRevA.76.042319} {\bibfield  {journal} {\bibinfo
  {journal} {Phys. Rev. A}\ }\textbf {\bibinfo {volume} {76}},\ \bibinfo
  {pages} {042319} (\bibinfo {year} {2007})}\BibitemShut {NoStop}%
\bibitem [{\citenamefont {Blais}\ \emph {et~al.}(2004)\citenamefont {Blais},
  \citenamefont {Huang}, \citenamefont {Wallraff}, \citenamefont {Girvin},\
  and\ \citenamefont {Schoelkopf}}]{blais2004circuitqed}%
  \BibitemOpen
  \bibfield  {author} {\bibinfo {author} {\bibfnamefont {A.}~\bibnamefont
  {Blais}}, \bibinfo {author} {\bibfnamefont {R.-S.}\ \bibnamefont {Huang}},
  \bibinfo {author} {\bibfnamefont {A.}~\bibnamefont {Wallraff}}, \bibinfo
  {author} {\bibfnamefont {S.~M.}\ \bibnamefont {Girvin}}, \ and\ \bibinfo
  {author} {\bibfnamefont {R.~J.}\ \bibnamefont {Schoelkopf}},\ }\bibfield
  {title} {\enquote {\bibinfo {title} {Cavity quantum electrodynamics for
  superconducting electrical circuits: An architecture for quantum
  computation},}\ }\href {\doibase 10.1103/PhysRevA.69.062320} {\bibfield
  {journal} {\bibinfo  {journal} {Phys. Rev. A}\ }\textbf {\bibinfo {volume}
  {69}},\ \bibinfo {pages} {062320} (\bibinfo {year} {2004})}\BibitemShut
  {NoStop}%
\bibitem [{\citenamefont {{De Raedt}}(1987)}]{deraedt1987productformula}%
  \BibitemOpen
  \bibfield  {author} {\bibinfo {author} {\bibfnamefont {H.}~\bibnamefont {{De
  Raedt}}},\ }\bibfield  {title} {\enquote {\bibinfo {title} {{Product formula
  algorithms for solving the time dependent Schr{\"o}dinger equation}},}\
  }\href {\doibase 10.1016/0167-7977(87)90002-5} {\bibfield  {journal}
  {\bibinfo  {journal} {Comput. Phys. Rep.}\ }\textbf {\bibinfo {volume} {7}},\
  \bibinfo {pages} {1} (\bibinfo {year} {1987})}\BibitemShut {NoStop}%
\bibitem [{\citenamefont {{De Raedt}}\ and\ \citenamefont
  {Michielsen}(2006)}]{deraedt2004computational}%
  \BibitemOpen
  \bibfield  {author} {\bibinfo {author} {\bibfnamefont {H.}~\bibnamefont {{De
  Raedt}}}\ and\ \bibinfo {author} {\bibfnamefont {K.}~\bibnamefont
  {Michielsen}},\ }\bibfield  {title} {\enquote {\bibinfo {title}
  {Computational methods for simulating quantum computers},}\ }in\ \href
  {http://arxiv.org/abs/quant-ph/0406210} {\emph {\bibinfo {booktitle} {Quantum
  and Molecular Computing, Quantum Simulations}}},\ \bibinfo {series} {Handbook
  of Theoretical and Computational Nanotechnology}, Vol.~\bibinfo {volume}
  {3},\ \bibinfo {editor} {edited by\ \bibinfo {editor} {\bibfnamefont
  {M.}~\bibnamefont {Rieth}}\ and\ \bibinfo {editor} {\bibfnamefont
  {W.}~\bibnamefont {Schommers}}}\ (\bibinfo  {publisher} {American
  Scientific},\ \bibinfo {address} {Los Angeles},\ \bibinfo {year} {2006})\
  pp.\ \bibinfo {pages} {1--48}\BibitemShut {NoStop}%
\bibitem [{\citenamefont {McKay}\ \emph {et~al.}(2017)\citenamefont {McKay},
  \citenamefont {Wood}, \citenamefont {Sheldon}, \citenamefont {Chow},\ and\
  \citenamefont {Gambetta}}]{McKay2016VZgate}%
  \BibitemOpen
  \bibfield  {author} {\bibinfo {author} {\bibfnamefont {D.~C.}\ \bibnamefont
  {McKay}}, \bibinfo {author} {\bibfnamefont {C.~J.}\ \bibnamefont {Wood}},
  \bibinfo {author} {\bibfnamefont {S.}~\bibnamefont {Sheldon}}, \bibinfo
  {author} {\bibfnamefont {J.~M.}\ \bibnamefont {Chow}}, \ and\ \bibinfo
  {author} {\bibfnamefont {J.~M.}\ \bibnamefont {Gambetta}},\ }\bibfield
  {title} {\enquote {\bibinfo {title} {Efficient $z$ gates for quantum
  computing},}\ }\href {\doibase 10.1103/PhysRevA.96.022330} {\bibfield
  {journal} {\bibinfo  {journal} {Phys. Rev. A}\ }\textbf {\bibinfo {volume}
  {96}},\ \bibinfo {pages} {022330} (\bibinfo {year} {2017})}\BibitemShut
  {NoStop}%
\bibitem [{\citenamefont {Gambetta}\ \emph {et~al.}(2011)\citenamefont
  {Gambetta}, \citenamefont {Motzoi}, \citenamefont {Merkel},\ and\
  \citenamefont {Wilhelm}}]{gambetta2010dragtheory}%
  \BibitemOpen
  \bibfield  {author} {\bibinfo {author} {\bibfnamefont {J.~M.}\ \bibnamefont
  {Gambetta}}, \bibinfo {author} {\bibfnamefont {F.}~\bibnamefont {Motzoi}},
  \bibinfo {author} {\bibfnamefont {S.~T.}\ \bibnamefont {Merkel}}, \ and\
  \bibinfo {author} {\bibfnamefont {F.~K.}\ \bibnamefont {Wilhelm}},\
  }\bibfield  {title} {\enquote {\bibinfo {title} {Analytic control methods for
  high-fidelity unitary operations in a weakly nonlinear oscillator},}\ }\href
  {\doibase 10.1103/PhysRevA.83.012308} {\bibfield  {journal} {\bibinfo
  {journal} {Phys. Rev. A}\ }\textbf {\bibinfo {volume} {83}},\ \bibinfo
  {pages} {012308} (\bibinfo {year} {2011})}\BibitemShut {NoStop}%
\bibitem [{\citenamefont {Gambetta}\ \emph {et~al.}(2012)\citenamefont
  {Gambetta}, \citenamefont {C\'orcoles}, \citenamefont {Merkel}, \citenamefont
  {Johnson}, \citenamefont {Smolin}, \citenamefont {Chow}, \citenamefont
  {Ryan}, \citenamefont {Rigetti}, \citenamefont {Poletto}, \citenamefont
  {Ohki}, \citenamefont {Ketchen},\ and\ \citenamefont
  {Steffen}}]{gambetta2012crosstalkSimRB}%
  \BibitemOpen
  \bibfield  {author} {\bibinfo {author} {\bibfnamefont {J.~M.}\ \bibnamefont
  {Gambetta}}, \bibinfo {author} {\bibfnamefont {A.~D.}\ \bibnamefont
  {C\'orcoles}}, \bibinfo {author} {\bibfnamefont {S.~T.}\ \bibnamefont
  {Merkel}}, \bibinfo {author} {\bibfnamefont {B.~R.}\ \bibnamefont {Johnson}},
  \bibinfo {author} {\bibfnamefont {J.~A.}\ \bibnamefont {Smolin}}, \bibinfo
  {author} {\bibfnamefont {J.~M.}\ \bibnamefont {Chow}}, \bibinfo {author}
  {\bibfnamefont {C.~A.}\ \bibnamefont {Ryan}}, \bibinfo {author}
  {\bibfnamefont {C.}~\bibnamefont {Rigetti}}, \bibinfo {author} {\bibfnamefont
  {S.}~\bibnamefont {Poletto}}, \bibinfo {author} {\bibfnamefont {T.~A.}\
  \bibnamefont {Ohki}}, \bibinfo {author} {\bibfnamefont {M.~B.}\ \bibnamefont
  {Ketchen}}, \ and\ \bibinfo {author} {\bibfnamefont {M.}~\bibnamefont
  {Steffen}},\ }\bibfield  {title} {\enquote {\bibinfo {title}
  {Characterization of addressability by simultaneous randomized
  benchmarking},}\ }\href {\doibase 10.1103/PhysRevLett.109.240504} {\bibfield
  {journal} {\bibinfo  {journal} {Phys. Rev. Lett.}\ }\textbf {\bibinfo
  {volume} {109}},\ \bibinfo {pages} {240504} (\bibinfo {year}
  {2012})}\BibitemShut {NoStop}%
\bibitem [{\citenamefont {Stephan}\ and\ \citenamefont
  {Doctor}(2015)}]{JUQUEEN}%
  \BibitemOpen
  \bibfield  {author} {\bibinfo {author} {\bibfnamefont {M.}~\bibnamefont
  {Stephan}}\ and\ \bibinfo {author} {\bibfnamefont {J.}~\bibnamefont
  {Doctor}},\ }\bibfield  {title} {\enquote {\bibinfo {title} {{JUQUUEN: IBM
  Blue Gene/Q Supercomputer System at the J{\"u}lich Supercomputing Centre}},}\
  }\href@noop {} {\bibfield  {journal} {\bibinfo  {journal} {J. of Large-Scale
  Res. Facil.}\ }\textbf {\bibinfo {volume} {1}},\ \bibinfo {pages} {A1}
  (\bibinfo {year} {2015})}\BibitemShut {NoStop}%
\bibitem [{\citenamefont {Gambetta}(2013)}]{gambetta2013controlIFF}%
  \BibitemOpen
  \bibfield  {author} {\bibinfo {author} {\bibfnamefont {J.~M.}\ \bibnamefont
  {Gambetta}},\ }\bibfield  {title} {\enquote {\bibinfo {title} {Control of
  superconducting qubits},}\ }in\ \href
  {https://books.google.de/books?id=-ms1nwEACAAJ} {\emph {\bibinfo {booktitle}
  {Quantum Information Processing: Lecture Notes of the 44th IFF Spring School,
  Schriften des Forschungszentrums J{\"u}lich, Reihe Schl{\"u}sseltechnologien
  / Key Technologies}}},\ Vol.~\bibinfo {volume} {52},\ \bibinfo {editor}
  {edited by\ \bibinfo {editor} {\bibfnamefont {D.~P.}\ \bibnamefont
  {DiVincenzo}}}\ (\bibinfo  {publisher} {Forschungszentrum J{\"u}lich},\
  \bibinfo {address} {Germany},\ \bibinfo {year} {2013})\BibitemShut {NoStop}%
\bibitem [{\citenamefont {Motzoi}\ \emph {et~al.}(2009)\citenamefont {Motzoi},
  \citenamefont {Gambetta}, \citenamefont {Rebentrost},\ and\ \citenamefont
  {Wilhelm}}]{motzoi2009drag}%
  \BibitemOpen
  \bibfield  {author} {\bibinfo {author} {\bibfnamefont {F.}~\bibnamefont
  {Motzoi}}, \bibinfo {author} {\bibfnamefont {J.~M.}\ \bibnamefont
  {Gambetta}}, \bibinfo {author} {\bibfnamefont {P.}~\bibnamefont
  {Rebentrost}}, \ and\ \bibinfo {author} {\bibfnamefont {F.~K.}\ \bibnamefont
  {Wilhelm}},\ }\bibfield  {title} {\enquote {\bibinfo {title} {Simple pulses
  for elimination of leakage in weakly nonlinear qubits},}\ }\href {\doibase
  10.1103/PhysRevLett.103.110501} {\bibfield  {journal} {\bibinfo  {journal}
  {Phys. Rev. Lett.}\ }\textbf {\bibinfo {volume} {103}},\ \bibinfo {pages}
  {110501} (\bibinfo {year} {2009})}\BibitemShut {NoStop}%
\bibitem [{\citenamefont {Nelder}\ and\ \citenamefont
  {Mead}(1965)}]{NelderMead1965}%
  \BibitemOpen
  \bibfield  {author} {\bibinfo {author} {\bibfnamefont {J.~A.}\ \bibnamefont
  {Nelder}}\ and\ \bibinfo {author} {\bibfnamefont {R.}~\bibnamefont {Mead}},\
  }\bibfield  {title} {\enquote {\bibinfo {title} {A simplex method for
  function minimization},}\ }\href {\doibase 10.1093/comjnl/7.4.308} {\bibfield
   {journal} {\bibinfo  {journal} {Comput. J.}\ }\textbf {\bibinfo {volume}
  {7}},\ \bibinfo {pages} {308} (\bibinfo {year} {1965})}\BibitemShut {NoStop}%
\bibitem [{\citenamefont {Nielsen}(2002)}]{nielsen2002gatefidelity}%
  \BibitemOpen
  \bibfield  {author} {\bibinfo {author} {\bibfnamefont {M.~A.}\ \bibnamefont
  {Nielsen}},\ }\bibfield  {title} {\enquote {\bibinfo {title} {A simple
  formula for the average gate fidelity of a quantum dynamical operation},}\
  }\href {\doibase 10.1016/S0375-9601(02)01272-0} {\bibfield  {journal}
  {\bibinfo  {journal} {Phys. Lett. A}\ }\textbf {\bibinfo {volume} {303}},\
  \bibinfo {pages} {249 } (\bibinfo {year} {2002})}\BibitemShut {NoStop}%
\bibitem [{\citenamefont {Wallman}\ \emph {et~al.}(2015)\citenamefont
  {Wallman}, \citenamefont {Granade}, \citenamefont {Harper},\ and\
  \citenamefont {Flammia}}]{Wallman2015unitarity}%
  \BibitemOpen
  \bibfield  {author} {\bibinfo {author} {\bibfnamefont {J.}~\bibnamefont
  {Wallman}}, \bibinfo {author} {\bibfnamefont {C.}~\bibnamefont {Granade}},
  \bibinfo {author} {\bibfnamefont {R.}~\bibnamefont {Harper}}, \ and\ \bibinfo
  {author} {\bibfnamefont {S.~T.}\ \bibnamefont {Flammia}},\ }\bibfield
  {title} {\enquote {\bibinfo {title} {Estimating the coherence of noise},}\
  }\href {http://stacks.iop.org/1367-2630/17/i=11/a=113020} {\bibfield
  {journal} {\bibinfo  {journal} {New J. Phys.}\ }\textbf {\bibinfo {volume}
  {17}},\ \bibinfo {pages} {113020} (\bibinfo {year} {2015})}\BibitemShut
  {NoStop}%
\end{thebibliography}%

\end{document}